\documentclass[aps,rmp,reprint,twocolumn,superscriptaddress]{revtex4-2}
\usepackage{enumerate,appendix,multirow,amssymb,amsfonts,dsfont,amsmath}
\usepackage{commath,braket}
\usepackage{color,calc,graphicx}
\usepackage[usenames,dvipsnames,svgnames,table,cmyk,hyperref]{xcolor}
\usepackage[charter,cal=cmcal,sfscaled=false]{mathdesign}
\definecolor{oeawblue}{cmyk}{0.9,0.68,0,0}
\definecolor{iqoqiblue}{cmyk}{0.76,0.11,0,0}
\definecolor{iffsred}{cmyk}{0.12,0.94,0.87,0.34}
\definecolor{thpurple}{cmyk}{0.65,1.0,0.0,0.2}
\definecolor{uestcblue}{cmyk}{0.99,0.78,0.16,0.03}

\newtheorem{thm}{\bf Theorem}
\newtheorem{clm}{\bf Claim} 
\newtheorem{cory}{\bf Corollary}

\usepackage{hyperref}
\usepackage{subfigure} 
\usepackage{lipsum} 
\usepackage[section]{placeins}

\hypersetup{
  pdftitle={Causality and Duality in Multipartite Generalized Probabilistic Theories},
  pdfauthor={Chen, Wang and Wang},
  pdfstartview=Fit,
  pdfpagelayout=SinglePage,
  colorlinks,
  linkcolor=uestcblue,
  citecolor=uestcblue,
  urlcolor=iffsred}
\usepackage[figure]{hypcap}

\begin{document}
\title{Causality and Duality in Multipartite  Generalized Probabilistic Theories}

\author{Yiying Chen}
\author{Peidong Wang}
\author{Zizhu Wang}
\affiliation{Institute of Fundamental and Frontier Sciences, University of Electronic Science and Technology of China, 611731, Chengdu, China}
\affiliation{Key Laboratory of Quantum Physics and Photonic Quantum Information, Ministry of Education, University of Electronic Science and Technology of China, 611731, Chengdu, China}

\begin{abstract}
Causality is one of the most fundamental notions in physics. Generalized probabilistic theories (GPTs) and the process matrix framework incorporate it in different forms. However, a direct connection between these frameworks remains unexplored. By demonstrating the duality between no-signaling principle and classical processes in tripartite classical systems, and extending some results to multipartite systems, we first establish a strong link between these two frameworks, which are two sides of the same coin. This provides an axiomatic approach to describe the measurement space within both box world and local theories. Furthermore, we describe a logically consistent 4-partite classical process acting as an extension of the quantum switch. By incorporating more than two control states, it allows both parallel and serial application of operations. We also provide a device-independent certification of its quantum variant in the form of an inequality.
\end{abstract}
\maketitle
\tableofcontents

\section{Introduction}
Causality serves as a foundational principle in physics which permeates its various branches. In general relativity, it can be used to reconstruct the topology of spacetime, up to a conformal factor~\cite{Malament,HawkingKingMcCarthy}. In our search for a deeper physical intuition behind quantum theory, causality also served as the seed for recent attempts to reconstruct it from information-theoretic principles~\cite{DAriano2017,navascuesGlanceQuantumModel2010,navascuesMacroscopicLocality2016,al-safiInformationCausalityEntropic2011,pawlowskiInformationCausalityPhysical2009,gallegoQuantumCorrelationsRequire2011,cabelloSimpleExplanationQuantum2013,fritzLocalOrthogonalityMultipartite2013,sainzExploringLocalOrthogonality2014,cabelloExclusivityPrincipleQuantum2014,NavascuesBoundingSetQuantum2007,Navascuesconvergenthierarchysemidefinite2008,navascuesAlmostQuantumCorrelations2015,sainzAlmostQuantumCorrelationsViolate2018}. Aiming to provide a unified framework that extends beyond classical and quantum theories, generalized probabilistic theories (GPTs) allow for the exploration of alternative causal structures and accommodating nonlocal phenomena within a consistent probabilistic model~\cite{barrettInformationProcessingGeneralized2007a,hardyQuantumTheoryFive2001a,mullerProbabilisticTheoriesReconstructions2021}.

Also motivated by causality, a different line of research on non-classical causal orders has greatly advanced our understanding of causality when the assumption of a pre-existing global causal order is dropped~\cite{oreshkovQuantumCorrelationsNo2012,baumelerSpaceLogicallyConsistent2016,BruknerQuantumcausality2014}. When only local consistency is enforced, the global causal order can be characterized by a process matrix~\cite{oreshkovQuantumCorrelationsNo2012} if the local theory is quantum, or by a classical process~\cite{baumelerSpaceLogicallyConsistent2016} if the local theory is classical. In both scenarios, global causal nonseperability can occur~\cite{baumelerMaximalIncompatibilityLocally2014,BaumelerPerfectsignalingthree2014d,AraujoWitnessingcausalnonseparability2015,Branciardsimplestcausalinequalities2015b}.

The understanding of causality in quantum foundations has developed from Bell locality, the no-signaling principle to causally non-separable processes. While previous researches have shown the relationships between different causal orders~\cite{leQuantumCorrelationsMinimal2023,fritzPolyhedralDualityBell2012,eftaxiasMultisystemMeasurementsGeneralized2023,SakharwadediagrammaticlanguageCausaloid2024a,liu2024tsirelson}, a direct connection between GPTs and the process matrix framework remains to be established.

This work focuses on two manifestations: the no-signaling principle and classical processes. We uncover a duality between these two principles in $(3,2,2)$ scenarios, where a tripartite system has two measurement choices per party, each with two possible outcomes. This finding establishes an explicit connection between box world and the process matrix framework. Moreover, connections to Bell locality naturally arise, further enriching the relationship between these frameworks. Specifically, we demonstrate that in the $(n,2,2)$ scenario, the spanning vectors of the measurement space in both box world and local theory are fully characterized by classical processes. This insight enables the formulation of a physical principle that governs measurements in both box world and local theory.

However, not all measurements within these frameworks are necessarily physically realizable. By demonstrating that certain effects stemming from probabilistic spanning vectors in measurement space violate the definitions within GPTs, we provide formal confirmation of the argument proposed by Baumeler and Wolf. Their work suggests that any modification of causal weights within the decomposition of a probabilistic process leads to a logical inconsistency~\cite{baumelerSpaceLogicallyConsistent2016}. This result highlights the limitations of allowing arbitrary modifications in the causal structure of probabilistic processes.

As a consequence of these findings, discussions of measurements in 4-partite systems must be limited to deterministic classical processes. By focusing on the valid measurements, we identify a novel causal order that governs the signaling channels, allowing them to operate either in parallel or serially. Additionally, we introduce a device-independent certification method for the quantum variant of this causal switch.

\section{Causality in the state space}
Causality, as a fundamental principle in physics, offers a comprehensive framework to describe the relationships between events, specifically through two key components: the state space and the measurement space. The state space represents the possible correlations between inputs and outputs within a joint system, constrained by the underlying physical laws.

Using the language of probability, the framework of generalized probabilistic theories (GPTs) has been developed to enable a fair comparison of physical theories~\cite{barrettInformationProcessingGeneralized2007a,hardyQuantumTheoryFive2001a,mullerProbabilisticTheoriesReconstructions2021}. In a GPT, denoted by $\mathcal{G}$, the joint state $\boldsymbol{P}_{\mathcal{G}}$ of an $n$-partite system is described as a list of probabilities $P_{\mathcal{G}}(\vec{a}|\vec{x})$, where $\vec{a} = \{a_1, \dots, a_n\}$ represents the outcomes given all possible fiducial measurements $\vec{x} = \{x_1, \dots, x_n\}$. The state space, denoted by $\Omega_{\mathcal{G}}$, is finite-dimensional and convex, consisting of all possible states allowed by the physical theory. Any valid state $\boldsymbol{P}_{\mathcal{G}} \in \Omega_{\mathcal{G}}$ must satisfy several conditions based on the properties of probability and basic physical assumptions:
\begin{align}
        \sum_{a_i} P_{\mathcal{G}}(\vec{a}|\vec{x}) \text{ is independent} &\text{ of } x_i, \forall\vec{x}\in \left\lbrace 1,\dots,n \right\rbrace  \label{eq:state_ns}
        \\
        \sum_{\boldsymbol{a }}P_{\mathcal{G}}(\vec{a}|\vec{x}) &=1, \forall \vec{x} 
        \label{eq:state_nor}
        \\
        P_{\mathcal{G}}(\vec{a}|\vec{x}) &\geq 0, \forall \vec{a},\vec{x}  \label{eq:state_nonne}
\end{align}

A fundamental causal principle derived from relativistic causality is the no-signaling principle, denoted by $\mathcal{NS}$~\cite{popescuQuantumNonlocalityAxiom1994a}. This principle ensures that information cannot travel faster than the speed of light, thereby constraining the correlations in physical systems, such as in the box world, to prevent superluminal signaling. Another key causality model within GPTs is local hidden variable theory, denoted by $\mathcal{L}$~\cite{bohmSuggestedInterpretationQuantum1952b,bohmSuggestedInterpretationQuantum1952d,belltEinsteinPodolskyRosen,clauserProposedExperimentTest1969}. In this model, a local hidden variable $\lambda$ carries all the information necessary to predict measurement outcomes. The theory imposes constraints on the states $\boldsymbol{P}_{\mathcal{L}}(\vec{a}|\vec{x})$, where local correlations are expressed as $P_{\mathcal{L}}(\vec{a}|\vec{x}) = \int_{\Lambda} q(\lambda) P(a_1|x_1, \lambda) \dots P(a_n|x_n, \lambda) d\lambda$, with $\vec{a}, \vec{x} \in \{0, 1\}$ and $\lambda$ having a well-defined probability distribution $q(\lambda)$, independent of the measurement settings $\vec{x}$. This theory thus provides a global causal order, where measurement outcomes are determined in a local and deterministic manner.

Geometrically, the state spaces in both box world and local hidden variable theory can be represented as polytopes~\cite{gunterm.zieglerLecturesPolytopes1995}. These polytopes are described in two equivalent ways: either as a bounded intersection of finitely many half-spaces or as a convex hull of finitely many vertices. This geometric interpretation allows for a deeper understanding of the structure of these state spaces and their corresponding constraints.

In a different approach to studying causality, the framework of classical processes offers a new perspective~\cite{baumelerSpaceLogicallyConsistent2016}. Classical processes, denoted by $\mathcal{CP}$, abandon the assumption of a global space-time structure while keeping classical probability theory locally valid~\cite{baumelerSpaceLogicallyConsistent2016}. This framework is based on the following assumptions: (1) Free randomness, meaning that the inputs to the systems are chosen freely; (2) Closed laboratories, where correlations between parties can only arise if they are causally connected; and (3) Locally valid classical systems, where local operations are restricted to classical, but there are no theoretical constraints on how boxes are related to each other causally.
\begin{figure}
    \centering
    \includegraphics[scale=0.33 ]{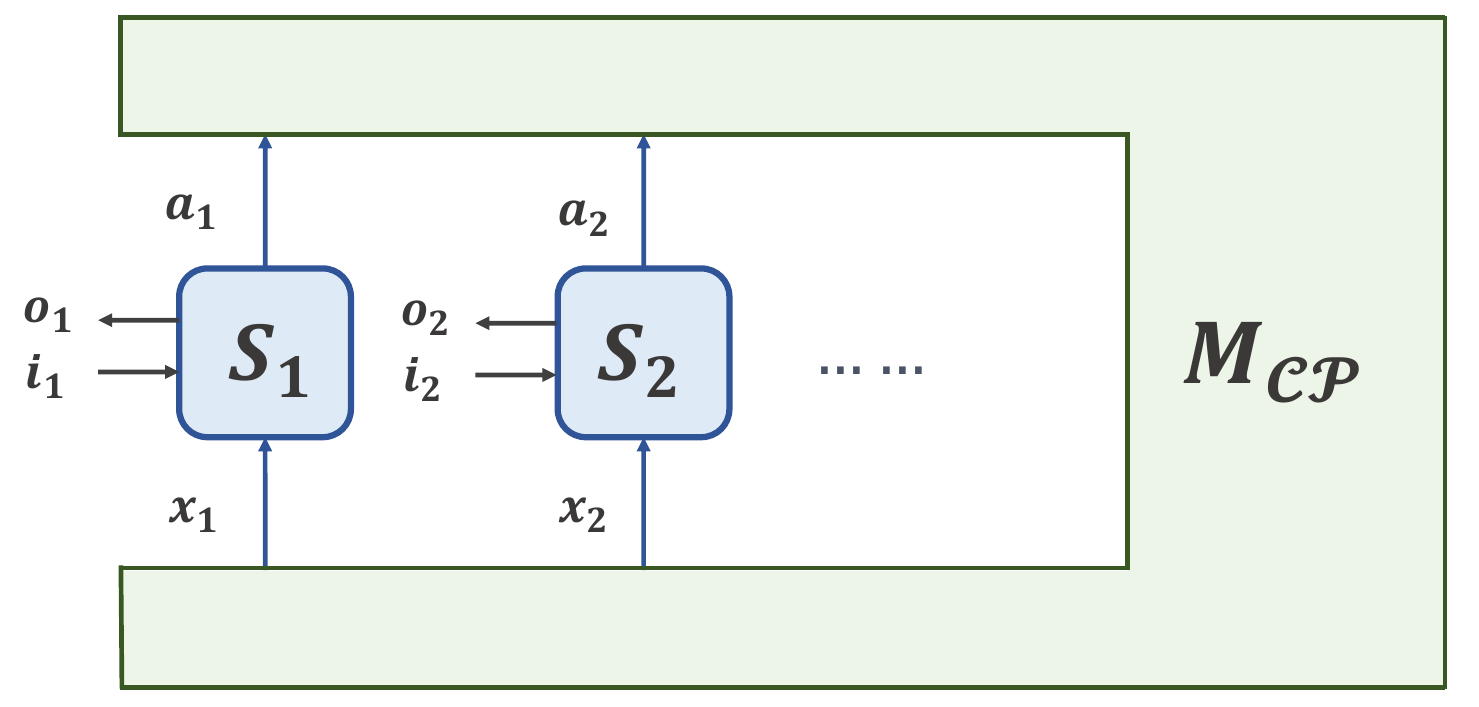}
    \caption{The framework of classical processes without predefined causal order. The local operation inside a box $S_j (j \in \{1,\dots,n \})$ shows that the output $a_j \in \mathcal{A}$ is a deterministic function of an input $x_j \in \mathcal{X}$. And the causal order is expressed as a logically consistent mapping $\boldsymbol{M}_{\mathcal{CP}}$ such that $\boldsymbol{M}_{\mathcal{CP}} : \mathcal{A} \to \mathcal{X}$. }
    \label{fig:system}
\end{figure}

In a $n$-partite system, the laboratory $S_j (j \in \lbrace 1,2,\dots, n \rbrace)$ is regarded as a black box with an free input variable $i_j$  and an output variable $o_j$, as shown in Fig.  \ref{fig:system}. $S_j$ get an input from the environment and send an output back, denoted by ${x_j} \in \mathcal{X}$ and ${a_j} \in \mathcal{A}$ respectively (the sets $\mathcal{X}$ and $\mathcal{A}$ are binary). The state of a single party $S_j$ is defined as a stochastic process $\boldsymbol{P}(a_j,o_j|x_j,i_j)$.  To simplify, we fix the value of $\{i_1,i_2,\dots,i_n\}$ and sum over $\{o_1,o_2,\dots,o_n\}$, i.e., $\sum_{\vec{o}} P(\vec{a},\vec{o}|\vec{x},\vec{i}=\vec{I})=P(\vec{a}|\vec{x})$.

\begin{figure}
    \centering
    \subfigure[]{\includegraphics[scale=0.55]{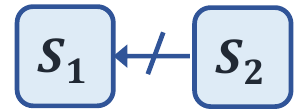}}
    \subfigure[]{\includegraphics[scale=0.55]{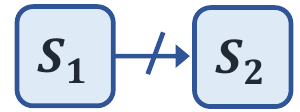}}
    \caption{Two possible causal orders on a bipartite system.  In (a), the output of $S_2$  is causally influenced by $S_1$, i.e., $P_1^{S_1 \nsucceq S_2}=P(a_1|x_1)P(a_2|a_1,x_1,x_2)$. In (b), the output of $S_1$ depends on $S_2$, i.e., $P_2^{S_2 \nsucceq S_1}=P(a_2|x_2)P(a_1|a_2,x_2,x_1)$. A predefined causal order is described as a convex combination of case (a) and (b).}
    \label{fig:party2}
\end{figure}

For each party $S_j$, $a_j$ causally depends on $x_j$, and $x_j$ is given by the environment. $x_j$ thus is in the causal past of $a_j$(denoted by $x_j \preceq a_j$), or equivalently, $a_j$ is in the causal future of $x_j$(denoted by $a_j \succeq x_j$). Moreover, if the output $a_k$ of party $S_k$ is correlated with the input $x_t$ of party $S_t$, we say that $S_t$ signals to $S_k$, i.e., $S_t \preceq S_k$ or $S_k \succeq S_t$. By assuming unidirectional signaling, a predefined causal order of two parties $S_1,S_2$ is defined as a convex mixture of possible causal orders, such that the probability distribution $P(a_1,a_2|x_1,x_2)$ is written as
 \begin{equation}
 \begin{split}
          P(a_1,a_2|x_1,x_2)= q P_1^{S_1 \nsucceq S_2}+(1-q) 
          P_2^{S_2 \nsucceq S_1},
          q\geq 0
     \label{eq:preCO}
 \end{split}
 \end{equation}
where $P_1^{S_1 \nsucceq S_2}$ and $P_2^{S_2 \nsucceq S_1}$ correspond to two one-way signaling distributions shown in Fig.  \ref{fig:party2}. Intuitively, a necessary condition for a predefined causal order of $n$ parties is that in each distribution $P_k(\vec{a}|\vec{x})$, there is at least one party $S_j$ that is not in the causal future of any other party, that is $S_j \nsucceq^{k} S_i (\forall  i \neq j)$. 

In a classical system, the stochastic process $ P(\vec{a}|\vec{x}) $ inside a box represents a map from inputs $ \vec{x} $ to outputs $ \vec{a} $, denoted by $ \boldsymbol{D}_{\mathcal{CP}}: \mathcal{X} \to \mathcal{A} $ (see Fig.  \ref{fig:system}). For a joint system, the set of operations is defined as $ \{\boldsymbol{D}_{\mathcal{CP}}: \boldsymbol{D}_{\mathcal{CP}} = \boldsymbol{D}^{(t_1)}_{\mathcal{CP}} \otimes \dots \otimes \boldsymbol{D}^{(t_n)}_{\mathcal{CP}}, \forall t_1, \dots t_n\} $, where $ \boldsymbol{D}^{(t_i)}_{\mathcal{CP}} $ refers to a local operation conducted by the $ i $-th party. Considering the local validity of classical theory, the operations $ \boldsymbol{D}_{\mathcal{CP}}^{(t)} $ for each party are limited to a specific set of transformations: constant outputs (0 or 1), the identity operation, and the bit-flip operation (see Eq. (\ref{eq:O})).
\begin{equation}
	\begin{split}  
		&\boldsymbol{D}^{(0)}_{ \mathcal{CP} }= \begin{pmatrix}1 & 1 \\ 0 & 0\end{pmatrix}
  \quad \quad
		\boldsymbol{D}^{(1)}_{ \mathcal{CP} }= \begin{pmatrix}1 & 0 \\ 0 & 1\end{pmatrix}
  \\
		&\boldsymbol{D}^{(2)}_{ \mathcal{CP} }=	\begin{pmatrix}0 & 1 \\1 & 0\end{pmatrix}
  \quad \quad
		\boldsymbol{D}^{(3)}_{ \mathcal{CP} }= \begin{pmatrix}0 & 0 \\1 & 1\end{pmatrix} 
	\end{split}
	\label{eq:O}	
 \end{equation}

\section{Causality in the measurement space}

As a complementary aspect of the state space, causality within the measurement space describes the relationships between different measurements and how one subsystem can influence others. Our investigations reveal the existence of global causal nonseparability in the measurement space, highlighting the intricate dependencies between the various subsystems.

In GPTs, measurements are well defined as maps. Firstly, an element of the dual of state space, denoted by $\boldsymbol{e}_{\mathcal{G}} \in \Omega_{\mathcal{G}}^*$, is called as effect if it is expressed as a linear functional that map from a state $\boldsymbol{P}_{\mathcal{G}}$ to a probability, i.e., $0 \leq  \boldsymbol{e}_{\mathcal{G}} \cdot\boldsymbol{P}_{\mathcal{G}} \leq 1$, where the notation "$\cdot$" means the inner product~\cite{mullerProbabilisticTheoriesReconstructions2021}. Therefore, the measured probability of obtaining an outcome $r$ is written as $p_{\mathcal{G}}(r) = \boldsymbol{e}_{\mathcal{G},r} \cdot\boldsymbol{P}_{\mathcal{G}}$. In box world and local theory, any effect is equivalent to a nonnegative effect $\boldsymbol{e}_{\mathcal{G}}$, i.e., for all $\vec{a}$ and $\vec{x}$, $e_{\mathcal{G}}(\vec{a}|\vec{x}) \geq 0$ $(\mathcal{G} \in \{\mathcal{NS},\mathcal{L}\})$~\cite{barrettInformationProcessingGeneralized2007a}.  

Considering the linearity of probability, any $d$-outcomes measurement corresponds to a collection of effects $\lbrace \boldsymbol{e}_{\mathcal{G},j}\rbrace_{j=1}^{d}$. It is natural to define a set of total measurements~\cite{shortStrongNonlocalityTradeoff2010a}. For simplicity, no distinction is made between total measurement and measurement in the following discussion. In box world and local theory, the measurement space can be spanned by a set of vectors, denoted by $\{\boldsymbol{M}_{\mathcal{G}} \}$, which satisfy Eqs. (\ref{eq:measure_nor})-(\ref{eq:measure_nonne})~\cite{barrettInformationProcessingGeneralized2007a}. Much like the state space, the measurement spaces in box world and local theory are described as convex polytopes.
\begin{align}
    \boldsymbol{M}_{\mathcal{G}} \cdot \boldsymbol{P}_{\mathcal{G}} = \sum_{j=1}^{d}& \boldsymbol{e}_{\mathcal{G},j} (\vec{a}|\vec{x})\cdot \boldsymbol{P}_{\mathcal{G}} (\vec{a}|\vec{x})  =1, \forall \boldsymbol{P}_{\mathcal{G}} \in \Omega_{\mathcal{G}} \label{eq:measure_nor}
    \\
     M_{\mathcal{G}} (\vec{a}|\vec{x}) &\geq 0, \mathcal{G} \in \{\mathcal{NS},\mathcal{L}\} , \forall \vec{a},\vec{x}\label{eq:measure_nonne}
\end{align}

The study of measurement spaces in GPTs primarily addresses the causal dependence between measurement events on individual subsystems~\cite{shortStrongNonlocalityTradeoff2010a,allcockClosedSetsNonlocal2009,brunnerNonlocalityDistillationPostquantum2009}. A fundamental aspect of this is that in a joint system,  a measurement can be defined as basic measurement or wiring if the outcomes of one individual measurement can influence subsequent measurements. However, in certain cases~\cite{shortStrongNonlocalityTradeoff2010a}, the structure of causality in measurement spaces is less apparent, making it difficult to identify intuitive notions of cause and effect. This ambiguity can obscure the physical meaning behind certain measurement processes.

To gain a comprehensive understanding of the physical interpretation of measurements, we introduce the concept of logically consistent classical processes without a predefined causal order~\cite{baumelerSpaceLogicallyConsistent2016}. In a classical process, causality between $n$ boxes is defined as a stochastic process mapping from output space to input space, represented by $\boldsymbol{M}_{\mathcal{CP}} : \mathcal{A} \to \mathcal{X}$~\cite{baumelerSpaceLogicallyConsistent2016}. Each party $S_i$ sends the output $a_i$ to causal mapping $\boldsymbol{M}_{\mathcal{CP}}$, and subsequently receives an input $x_i$ from the same mapping (see Fig.  \ref{fig:system}). Rather than expressing this relationship as $\boldsymbol{M}_{\mathcal{CP}}(\vec{x}|\vec{a})$, we adopt the notation 
$\boldsymbol{M}_{\mathcal{CP}}(\vec{a}|\vec{x})$ for clarity.

On multi-partite systems, under any operations $\boldsymbol{D}_{ \mathcal{CP} }$, the probability that $n$ parties receiving $\vec{a}=\vec{r}$ from $\boldsymbol{M}_{\mathcal{CP}}$ is expressed as $p(\vec{r})=\sum_{\vec{x}} M_{\mathcal{CP}} (\vec{a}=\vec{r}| \vec{x})  D_{\mathcal{CP}}(\vec{a}=\vec{r}| \vec{x})$. To ensure the normalization and non-negativity of probability, the set of classical processes, denoted by $\{\boldsymbol{M}_{\mathcal{CP}} \}$, satisfies 
\begin{align}
 \begin{split}
     \boldsymbol{M}_{ \mathcal{CP} } \cdot \boldsymbol{D} 
 =\sum_{\vec{a},\vec{x}}M_{\mathcal{CP}}(\vec{a}|\vec{x})&D(\vec{a}| \vec{x})=1, \forall \boldsymbol{D} \in \{\boldsymbol{D}_{ \mathcal{CP} }\}
 \label{eq:E_nor}
 \end{split}
 \\
	M_{ \mathcal{CP} }(\vec{a}|\vec{x})&\ge 0, \forall \vec{a},\vec{x}		\label{eq:E_nonne}
 \end{align}

By Eqs. (\ref{eq:E_nor})-(\ref{eq:E_nonne}), the set of classical processes is also mathematically represented as a polytope. In this representation, the vertex set is roughly classified into two types: the set of extremal deterministic classical processes, denoted by $\{\boldsymbol{M}_{\mathcal{CP}}^{D}\}$, and the set of extremal classical processes where the elements are probabilistic. While, the latter are demonstrated to be invalid in the following section. The set $\{\boldsymbol{M}_{\mathcal{CP}}^{D}\}$ can be further divided into $n+1$ classes by the number of parties that receiving a constant input from the mapping. For instance, if there are $d$ deterministic classical processes involving $k$ parties $(k \leq n)$ receiving a constant value, we denote this subset by $\{\boldsymbol{M}_{\mathcal{CP},i}^{D(k)}\}_{i=1}^{d}$. 

Moreover, the causal orders among the parties are represented by Boolean functions, which can be illustrated using a truth table. In this framework, interactions between two parties are facilitated through an external single-bit channel embedded within $\boldsymbol{M}_{\mathcal{CP}}$.  This channel can either implement the identity operation or a bit-flip operation. As the number of parties increases, complex interactions are characterized by a multi-input Boolean function. An example of this is depicted in Fig.  ~\ref{fig:ns_cp_CP}.

\section{Physical principles characterizing states and measurements}
While Bell's locality, the no-signaling principle, and classical processes describe different causal orders in physical systems, the explicit relationship among them remains incomplete. Previous research has shown that the state space of local theory is dual to that of box world~\cite{fritzPolyhedralDualityBell2012, leQuantumCorrelationsMinimal2023}. Building on this, our results reveal an interesting symmetry: the measurement space in both local theory and box world have equivalent spanning vectors. Extending earlier efforts to connect GPTs with the process matrix framework~\cite{eftaxiasMultisystemMeasurementsGeneralized2023, SakharwadediagrammaticlanguageCausaloid2024a}, we demonstrate that the set of classical processes $\{\boldsymbol{M}_{\mathcal{CP}}\}$ forms the polar dual of the state space in box world in tripartite scenarios. Furthermore, we show that the spanning sets of measurement vectors for both box world and local theory can be physically characterized by classical processes.

To begin, we focus on box world and local theory. While these theories are well understood in terms of their state spaces, their dual spaces are less clearly defined. We find that the vector space of measurements in box world can be spanned by local measurement vectors, $\lbrace \boldsymbol{M}_{\mathcal{L}} \rbrace$. Using the fact that the no-signaling probability distribution forms an affine hull of the local probability distribution~\cite{abramskySheafTheoreticStructureNonLocality2011}, any no-signaling box can be expressed as an affine combination of local deterministic states. Considering Eqs. (\ref{eq:measure_nor})-(\ref{eq:measure_nonne}), this shows that the measurement spaces in both local theory and box world are spanned by the same set of vectors, i.e., $\lbrace \boldsymbol{M}_{\mathcal{NS}} \rbrace = \lbrace \boldsymbol{M}_{\mathcal{L}} \rbrace$.

The equivalence between the set of spanning measurement vectors in box world and those in local theory reveals connections between the dual spaces of distinct physical correlations. By further investigating scenarios within box world, we gain insights through the rich geometric structures of polytopes. A key property of such structures is that every non-empty $d$-polytope $P$ admits a polar dual polytope $P^*$, defined as $P^*=\{ y \in \mathbb{R}^d: x^{T}y \leq 1, \forall x \in P \}$~\cite{Fukuda,gunterm.zieglerLecturesPolytopes1995}. This duality provides a fundamental way for understanding and analyzing such systems, offering powerful tools to explore the connections between different physical theories.

\begin{thm}\!\!\textbf{.}\label{thm1}
     The set of classical processes, $\{\boldsymbol{M}_{\mathcal{CP}} \}$, is the polar dual of the state space in box world,  $\{ \boldsymbol{P}_{\mathcal{NS}} \}$, in $(2,2,2)$ and $(3,2,2)$ scenarios.
\end{thm}

In geometric terms, the duality can be more intuitively expressed as a correspondence between the vertices of the primal polytope and the facets of its dual polytope. In this case, it indicates that the faces defining the state space in box world, as expressed in Eqs. (\ref{eq:state_ns})-(\ref{eq:state_nonne}), are equivalent to a set of conditions represented by $\{ \boldsymbol{P}: \boldsymbol{M} \cdot \boldsymbol{P} = 1, \forall \boldsymbol{M} \in \{\boldsymbol{M}_{\mathcal{CP}} \}, \text{ and } P(\vec{a}|\vec{x}) \geq 0, \forall \vec{a}, \vec{x} \}$. The non-negativity conditions are naturally satisfied by the definition of spanning vectors in measurement space. Furthermore, we demonstrate that the set of equations $ \{ \boldsymbol{M} \cdot \boldsymbol{P} = 1, \forall \boldsymbol{M} \in \{\boldsymbol{M}_{\mathcal{CP}} \}\} $ implies the constraints given by Eqs. (\ref{eq:state_ns})-(\ref{eq:state_nor}) in box world for the $(2,2,2)$ and $(3,2,2)$ scenarios.

Firstly, we show that the set of equations $ \{ \boldsymbol{M} \cdot \boldsymbol{P} = 1, \forall \boldsymbol{M} \in \{ \boldsymbol{M}_{\mathcal{CP},j}^{D(n)} \}_{j=1}^{2^n} \} $ represents the normalization condition for states. Here, $ \boldsymbol{M}_{\mathcal{CP},j}^{D(n)} $ denotes the $ j $-th classical process where the inputs of $ n $ parties are fixed to a given constant values $ \vec{x} = \vec{X}$. Specifically, for each $j$, the equation $ \boldsymbol{M}_{\mathcal{CP},j}^{D(n)} \cdot \boldsymbol{P} = \sum_{\vec{a}} P(\vec{a}|\vec{x} = \vec{X}) = 1 $ ensures that  the total probability over all outcomes for a given input configuration sums to 1.

\begin{figure}[htbp]
	\centering
	\includegraphics[scale=0.23 ]{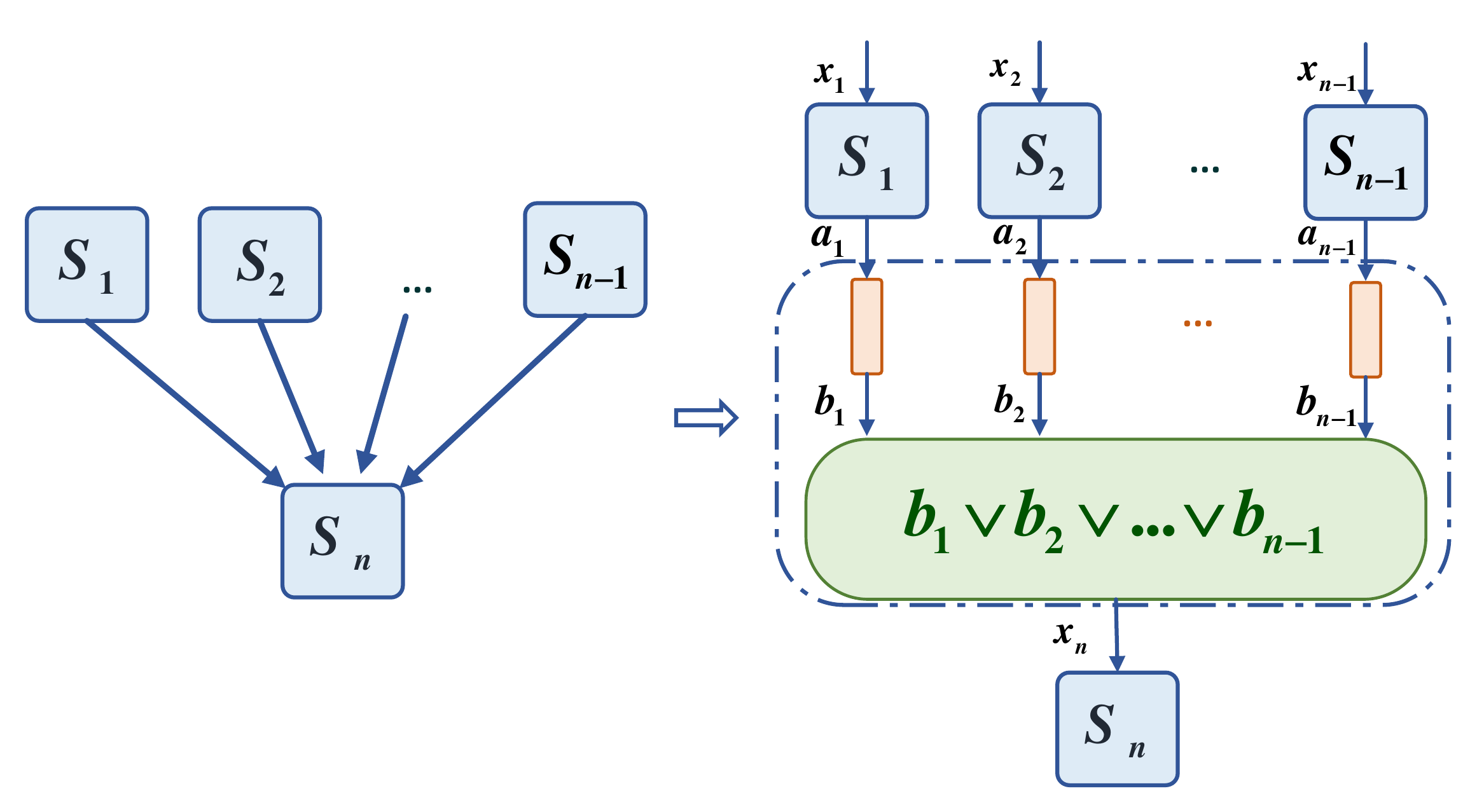}
	\caption{\label{fig:ns_cp_CP} An example of classical processes $\{ \boldsymbol{M}_{\mathcal{CP},i}^{D\left( n-1 \right) } \}_{i=1}^{ n\times 2^{2(n-1)}}$. The logically consistent map is describe by the dashed part. $\boldsymbol{M}_{\mathcal{CP},i}^{D\left( n-1 \right) }$ transforms each possible output $a_j (j = \left\lbrace 1,...,n-1 \right\rbrace )$ to $b_j$ through an identity or bit-flip channel (orange blocks). Subsequently, a $n-1$-to-$1$-bit Boolean function is applied (a green block), giving rise to $x_n=0$ only if $\{b_1,\dots,b_2\}=\{0,\dots,0\}$, otherwise $x_n=1$.  }
\end{figure}

Next, we derive the no-signaling principle from Gaussian elimination applied to the system of equations $\{ \boldsymbol{M} \cdot \boldsymbol{P}=1, \forall \boldsymbol{M} \in \{ \boldsymbol{M}_{\mathcal{CP},i}^{D\left( n-1 \right) } \}_{i=1}^{ n\times 2^{2(n-1)}}  \}$ in combination with $ \{ \boldsymbol{M} \cdot \boldsymbol{P} = 1, \forall \boldsymbol{M} \in \{ \boldsymbol{M}_{\mathcal{CP},j}^{D(n)} \}_{j=1}^{2^n} \} $  (see Eq. (\ref{eq:ns_cp_NS})). In this case, $\boldsymbol{M}_{\mathcal{CP},i}^{D\left( n-1 \right) }$  represents a  scenario where $n-1$ parties receive constant inputs, and one remaining party is influenced by all previous parties. Consider a classical process $\boldsymbol{M}_{\mathcal{CP},i}^{D(n-1)}$ involving a set of given single-bit channels, as shown in Fig.  \ref{fig:ns_cp_CP}. In this setup, the first $n-1$ parties, $S_1, \dots, S_{n-1}$, receive constant inputs $\{X_1, \dots, X_{n-1}\}$, while $S_n$ is in their causal future, meaning $S_i \nsucceq S_n$ for all $i \neq n$. Following each party $S_i$ acting as the causal past, the single-bit channel transforms the output $a_i$ into a new bit $b_i$ (for $i \in \{1, \dots, n-1\}$). When a Boolean function of the form $x_n = b_1 \vee \dots \vee b_{n-1}$ is applied, $S_n$ receives $x_n = 0$ only when all preceding bits $\{b_1, \dots, b_{n-1}\} = \{0, \dots, 0\}$, which corresponds to the initial output string $\{a_1, \dots, a_{n-1}\} = \{A_1, \dots, A_{n-1}\}$.  
\begin{equation}
	\begin{split}
		&\boldsymbol{M}_{\mathcal{CP},i}^{D\left( n-1 \right) } \cdot \boldsymbol{P}-
		\boldsymbol{M}_{\mathcal{CP},j}^{D\left( n \right) } \cdot \boldsymbol{P}
			\\
		&= \sum_{a_n} P (A_1,\dots,A_{n-1},a_n|X_1,\dots,X_{n-1},0) 
                \\
        &+ \sum_{  A'_1,\dots,A'_{n-1} ,a_n }  P (A'_1,\dots,A'_{n-1},a_n|X_1,\dots,X_{n-1},1) 
                \\
        &-\sum_{a_1,\dots,a_{n}} P (a_1,\dots,a_n|X_1,\dots,X_{n-1},1) 
			\\
		&= \sum_{a_n} P (A_1,\dots,A_{n-1},a_n|X_1,\dots,X_{n-1},0) 
                \\
        &-\sum_{a_n}  P(A_1,\dots,A_{n-1},a_n|X_1,\dots,X_{n-1},1)
			\\
		&=0
	\end{split} \label{eq:ns_cp_NS}
\end{equation} 
where $\left\{A'_1, \dots,A'_{n-1}\right\} \in\{0,1\}^{\otimes n-1}/$ $\left\{A_1,\dots, A_{n-1}\right\}$. By systematically enumerating all possible sets of fixed single-bit channels, relabeling constant input strings, and relabeling the party in the causal future, we derive a complete set of no-signaling conditions through Gaussian elimination.

We numerically demonstrate that for both bipartite and tripartite systems, given other extremal classical processes, including extremal probabilistic processes, the corresponding equations are entirely redundant. This results in a clear correspondence between the set of conditions $\{ \boldsymbol{M} \cdot \boldsymbol{P} = 1, \forall \boldsymbol{M} \in \{\boldsymbol{M}_{\mathcal{CP}}\} \}$ and Eqs. (\ref{eq:state_ns})–(\ref{eq:state_nor}).

In a similar manner, we show that the set $\{\boldsymbol{M}: \boldsymbol{M} \cdot \boldsymbol{P} = 1, \forall \boldsymbol{P} \in \Omega_{\mathcal{NS}}, \text{ and } M(\vec{a}|\vec{x}) \geq 0, \forall \vec{a}, \vec{x}\}$ translates directly into the constraints defining classical processes, as expressed by Eqs. (\ref{eq:E_nor})–(\ref{eq:E_nonne}). As previously discussed, the nonnegativity conditions are automatically satisfied due to the fundamental properties of probability theory. Finally, since the local operations in $\{\boldsymbol{D}_{\mathcal{CP}}\}$ imply local deterministic states (as established in Theorem \ref{thm2}), we verify that $\{\boldsymbol{M} \cdot \boldsymbol{P} = 1, \forall \boldsymbol{P} \in \Omega_{\mathcal{NS}}\}$ is equivalent to Eq. (\ref{eq:E_nor}).

In summary, we conclude that $\{\boldsymbol{M}_{\mathcal{CP}}\}$ is polar dual of $\Omega_{\mathcal{NS}}$ in the $(2,2,2)$ and $(3,2,2)$ scenarios. In the $(4,2,2)$ scenario, we reach similar conclusions, which will be discussed in detail later. However, extending the duality to the general $(n,2,2)$ case is challenging due to the complexity introduced by the vertex enumeration in a multipartite system.

Building on the established connections between local theories, box worlds, and classical processes, we observe that the measurement behaviors in both box worlds and local theories challenge conventional assumptions about global space-time structures~\cite{shortStrongNonlocalityTradeoff2010a}. We find that classical processes complement the more flexible causality observed in box worlds and local theories, revealing deeper links between classical and non-classical systems.

\begin{thm}\!\!\textbf{.}\label{thm2}
     Both the sets of spanning vectors for measurement space in box world and local theory, $\lbrace \boldsymbol{M}_{\mathcal{NS}} \rbrace$ and $\lbrace \boldsymbol{M}_{\mathcal{L}} \rbrace$, are fully characterized by the classical processes $\lbrace \boldsymbol{M}_{\mathcal{CP}} \rbrace$ in $(n,2,2)$ scenario $(n \geq 2)$.
\end{thm}

Given the equivalence of the sets of spanning vectors between box world and local theory, we will focus on the relationship between the sets $\{\boldsymbol{M}_{\mathcal{L}}\}$ and $\{\boldsymbol{M}_{\mathcal{CP}}\}$. Since that both sets are constrained to be non-negative, they will exhibit identical facets  if we can show that the equations governing the redundant variables are equivalent.

According to Fine's theorem~\cite{scaraniBellNonlocality2019}, each local deterministic state is compatible with a distinguishing rule where the outputs of the boxes are determined by a hidden variable, denoted by $\boldsymbol{Q}_{\mathcal{L}}(a_1,\dots,a_n,x_1,\dots,x_n)$. Then we show that each local deterministic state  corresponds to a local operation $\boldsymbol{D}_{\mathcal{CP}}$, in the context of n-partite scenarios:
 \begin{equation}
 \begin{split}  
    \boldsymbol{Q}_{\mathcal{L}}&(a_1,\dots,a_n,x_1,\dots,x_n) 
      =\boldsymbol{D}_{\mathcal{CP}}^{(t_1)} \otimes \dots \otimes \boldsymbol{D}_{\mathcal{CP}}^{(t_n)}
      \\
      &t_1,\dots, t_n \in \lbrace 0,1,2,3 \rbrace, \forall \boldsymbol{Q}_{\mathcal{L}} \in \Omega_{\mathcal{L}}
      \label{eq:ns_l}
 \end{split}
 \end{equation}
where the superscript $\{ t_1, \dots, t_n \}$ is represented by a string of binary bits, $\{ a_1 a_2, \dots, x_{n-1} x_n \}$. Consequently, the spanning vectors in local measurement space constrained by Eqs.~(\ref{eq:measure_nor})-(\ref{eq:measure_nonne}) are identified with classical processes satisfying Eqs.~(\ref{eq:E_nor})-(\ref{eq:E_nonne}). Therefore, we have $\{ \boldsymbol{M}_{\mathcal{NS}} \} = \{ \boldsymbol{M}_{\mathcal{L}} \} = \{ \boldsymbol{M}_{\mathcal{CP}} \}$. In the following section, we use the notation of classical processes to represent the identical sets of spanning vectors in box world and local theory.
\begin{figure}[htbp]
	\centering
	\includegraphics[scale=0.45]{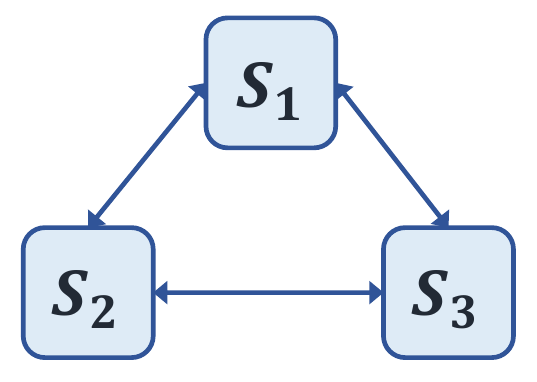}
	\caption{\label{fig:circle}  $\{ \boldsymbol{M}_{\mathcal{CP}}^{D\left( 0 \right) } \}$ is a set of self-circle causal orders, where it is not accessible by predefined causal order.}
\end{figure}

In summary, classical processes offer a pathway to explore and define physical principles that reveal the underlying structure of measurements. Importantly, these processes fits well with the previously established  concept of causality in GPTs~\cite{shortStrongNonlocalityTradeoff2010a,allcockClosedSetsNonlocal2009,brunnerNonlocalityDistillationPostquantum2009}. For instance, basic measurements in this framework often assume globally predefined causal orders~\cite{shortStrongNonlocalityTradeoff2010a}. In particular, non-causal classical processes play a key role in explaining measurements that standard GPTs cannot fully explain. For example, Short and Barrett give an example in their proofs, denoted by $\boldsymbol{M}_{\mathcal{CP}}^{D( 0) }$. It is a valid measurement that cannot be constructed from a simple probabilistic mixture of basic measurements (see Eq. (\ref{eq:circle}), where $\Bar{a}$ is the negation of $a$)~\cite{shortStrongNonlocalityTradeoff2010a}. Although incompatible with predefined global causal orders, this measurement fits within the framework of classical processes, representing an indefinite causal order involving directed cycles (see Fig\ref{fig:circle}).
\begin{equation}
	\boldsymbol{M}_{\mathcal{CP}}^{D( 0) }=\begin{bmatrix}
		0 & 0 & 0 & 0 & 1& 0 & 0 &0  \\
		1 & 0 & 0 & 0 & 0& 0 & 0 &0  \\
		0 & 1 & 0 & 0 & 0& 0 & 0 &0  \\
		0 & 1 & 0 & 0 & 0& 0 & 0 &0  \\
		0 & 0 & 0 & 0 & 1& 0 & 0 &0  \\
		0 & 0 & 1 & 0 & 0& 0 & 0 &0  \\
		1 & 0 & 0 & 0 & 0& 0 & 0 &0  \\
		0 & 0 & 1 & 0 & 0& 0 & 0 &0      
		\label{eq:circle}
	\end{bmatrix} 
\end{equation}
\begin{equation}
	x_1= \Bar{a}_2 \wedge \Bar{a}_3, \quad   x_2 = a_1 \wedge a_3, \quad  x_3 = \bar{a}_1 \wedge a_2 \label{eq:causal circle}
\end{equation}

\section{Box world and local theory require fine-tuning}
In classical causal models, statistical relationships between variables are expected to be robust to small parameter changes. When a model requires precise tuning, where any small variation leads to significant changes, this is called "fine-tuning". The study of fine-tuning is critical for assessing the sensitivity of causal models, which gains deeper insights into the nature of physical processes. In causal models that describe quantum systems, fine-tuning is unavoidable to reproduce statistical independencies, such as those in nonlocal correlations \cite{woodLessonCausalDiscovery2015a}. It highlights the limitations of causal models in explaining quantum behavior. Notably, fine-tuning is also observed in the measurement space of box world and local theories. Previous study has suggested that extremal probabilistic classical processes involve fine-tuned mixtures of deterministic points, with at least one point being logically inconsistent \cite{baumelerSpaceLogicallyConsistent2016}. Using the definition of effects in GPTs, We formally confirm this result in $(n,2,2)$ scenarios.

Since that physical rationality in a classical process implies determinism, we only discuss $\{0,1\}$-valued right stochastic matrix $\boldsymbol{Z}$, where $\sum_{\vec{x}} Z(\vec{a}|\vec{x})=1$ for all $\vec{a}$. To concisely describe the matrix, the information of $\boldsymbol{Z}$ is represented as a set of indices where $Z \left( \vec{a}| \vec{x} \right)=1$, denoted by $\mathds{1}(\boldsymbol{Z})$. The size of this set is represented by $|\mathds{1}(\boldsymbol{Z})|$. Note that each element index $( \vec{a}, \vec{x} )$ is a shorthand for $( a_1a_2\dots a_n,x_1x_2\dots x_n) $, when comparing two distinct inputs (outputs) $\vec{x}$ and $\vec{x}'$ (or $\vec{a}$ and $\vec{a}'$), we define $S_{id} (\vec{x},\vec{x}' )$ (or $S_{id} (\vec{a},\vec{a}' )$) as the index set of identical bits between $\vec{x}$ and $\vec{x}'$ (or between $\vec{a}$ and $\vec{a}'$), where $0 \leq |S_{id}| \leq n$. For instance, in a 6-partite system, given $\vec{a}=010010$ and $\vec{a}'=010001$, the index set of identical bits is represented as $S_{id} (\vec{a},\vec{a}')=\left\lbrace 1,2,3,4 \right\rbrace$.

\textbf{Definition 1.} A set of normal effects naturally emerge from $\{\boldsymbol{M}_{\mathcal{CP}}^{D} \} $ by changing any number of 1s to 0s for each $\boldsymbol{M}_{\mathcal{CP}}^{D}$, denoted by $E_{\mathcal{CP}} = \{ \boldsymbol{e}_{\mathcal{CP}}:\mathds{1} \left( \boldsymbol{e}_{\mathcal{CP}} \right) \subset \mathds{1} \left( \boldsymbol{Z} \right) \text{, and } 0 \leq |\mathds{1} \left( \boldsymbol{e}_{\mathcal{CP}} \right)|< 2^n , \forall \boldsymbol{Z} \in \{\boldsymbol{M}_{\mathcal{CP}}^{D} \} \}$.

Our interest lies in the physical interpretation of  $\{0,1\}$-valued effects beyond $E_{\mathcal{CP}}$.

\textbf{Definition 2.} A $\{0,1\}$-valued matrix is defined as an extra spanning measurement vectors, denoted by ${\boldsymbol{M}_{\mathcal{CP}}^{D^{\mathcal{C}}}}$, when it belongs to a complement of $\{ \boldsymbol{M}_{\mathcal{CP}}^{D}\}$. That is, $\{ \boldsymbol{M}_{\mathcal{CP}} \} \cup \{ {\boldsymbol{M}_{\mathcal{CP}}^{D^{\mathcal{C}}}} \}  = \{ \boldsymbol{Z}\in \mathds{R}^{2^n \times 2^n}| |\mathds{1} \left( \boldsymbol{Z} \right):= 2^n \text{, and }\sum_{\vec{x}} Z (\vec{a}|\vec{x})=1, \forall \vec{a} \}$.

\textbf{Definition 3.} A set of extra effects is obtained by changing any number of 1s to 0s for each ${\boldsymbol{M}_{\mathcal{CP}}^{D^{\mathcal{C}}}}$, denoted by $ E_{\mathcal{CP}}^{\mathcal{C}}=\{\boldsymbol{e}_{\mathcal{CP}}^{\mathcal{C}}|  \mathds{1} \left( \boldsymbol{e}_{\mathcal{CP}}^{\mathcal{C}} \right) \subset \mathds{1} \left( \boldsymbol{Z} \right) \text{, and } 0 \leq |\mathds{1} \left( \boldsymbol{e}_{\mathcal{CP}}^{\mathcal{C}} \right)|< 2^n , \forall \boldsymbol{Z} \in \{ {\boldsymbol{M}_{\mathcal{CP}}^{D^{\mathcal{C}}} } \} \}$. 

Note that a probabilistic classical processes is regarded as a proper mixture of measurements that includes at least one extra measurements. To understand the characteristics of extra effects and extra measurements, we firstly discuss the set of normal effects $E_{\mathcal{CP}}$.

\begin{clm}\!\!\textbf{.}\label{clm1}
    A $2^n \times 2^n$ $\{0,1\}$-valued matrix $\boldsymbol{Z}$, where $  0 \leq | \mathds{1} (\boldsymbol{Z}) | < 2^n$, sufficiently qualifies as a normal effect, if it satisfies one of the following 2 scenarios.

    Case 1. The number of nonzero elements is given by $| \mathds{1} (\boldsymbol{Z}) | =0$ or $| \mathds{1} (\boldsymbol{Z}) | =1$.

    Case 2. When the number of nonzero elements satisfies $| \mathds{1} (\boldsymbol{Z}) | \geq 2$, for any two distinct indices $( \vec{a},\vec{x} ),(\vec{a}', \vec{x}') \in \mathds{1}(\boldsymbol{Z})$,  there exists at least one index $q \in S_{id}(\vec{x},\vec{x}')$, where $S_{id}(\vec{x},\vec{x}') \subseteq \left\lbrace 1,\dots, n \right\rbrace(0 <|S_{id}| < n)$, such that $a_q \ne a'_q$.
\end{clm}

\textbf{Case 1.1 ($| \mathds{1} (\boldsymbol{Z}) | =0$)}:
The all-zero matrix is trivially derived from any total measurement by setting all non-zero elements to zero. 

\textbf{Case 1.2 ($| \mathds{1} (\boldsymbol{Z}) | =1$)}:
Given $\{\boldsymbol{M}^{D(n)}_{\mathcal{CP},i}\}_{i=1}^{2^n}$, where $n$ parties receive constant inputs, the sum of them leads to an all-one matrix. For example, Eq. (\ref{Appendeq:matrix}) represents a causal order where each party receives a constant input $0$.  Consequently, any $\{0,1\}$-valued matrix with $|\mathds{1}(\boldsymbol{Z})|=1$ is obtained by decomposing one  of the matrices from $\{\boldsymbol{M}^{D(n)}_{\mathcal{CP},i}\}_{i=1}^{2^n}$.
\begin{equation}
	\boldsymbol{M}_{\mathcal{CP},1}^{D(n)} = 
	\begin{bmatrix}
		1 & 0 & \dots & 0  \\
		1 & 0 & \dots & 0  \\
		\vdots &\vdots & \ddots & \vdots \\
		1 & 0 & \dots & 0
	\end{bmatrix} _{2^n \times 2^n}\label{Appendeq:matrix}
\end{equation}

\textbf{Case 2 ($| \mathds{1} (\boldsymbol{Z}) | \geq 2$)}:  
Using the rule of the inner product and the linearity of convex polytope,  Eq. (\ref{eq:E_nor}) implies that, if $\boldsymbol{Z}$ is a normal effect, we have $|\mathds{1} \left( \boldsymbol{Z} \right) \cap \mathds{1} \left( \boldsymbol{D}  \right) |\leq 1$, for all $\boldsymbol{D} \in \{\boldsymbol{D}_{ \mathcal{CP} ,i}\}_{i=1}^{n}$. The question regarding whether $\boldsymbol{Z}$ is a normal effect transforms into whether a set of joint operations $\{\boldsymbol{D}_{ \mathcal{CP} }\}$ satisfies 
\begin{equation}
    \begin{split}
		\{\boldsymbol{D}_{\mathcal{CP}}| ( \vec{a},\vec{x}) \in \mathds{1} \left(  \boldsymbol{D}_{\mathcal{CP}} \right) \} &\cap \{\boldsymbol{D}'_{\mathcal{CP}}| (\vec{a}', \vec{x}') \in \mathds{1} \left(  \boldsymbol{D}'_{\mathcal{CP}} \right) \}= \emptyset 
        \\
         \forall ( \vec{a},\vec{x}),(\vec{a}', \vec{x}') &\in \mathds{1}(\boldsymbol{Z}) ,  ( \vec{a},\vec{x}) \ne (\vec{a}', \vec{x}')
    \label{Appendeq:empty}	        
    \end{split}
\end{equation}

\begin{figure}[htbp]
	\centering
	\includegraphics[scale=0.28]{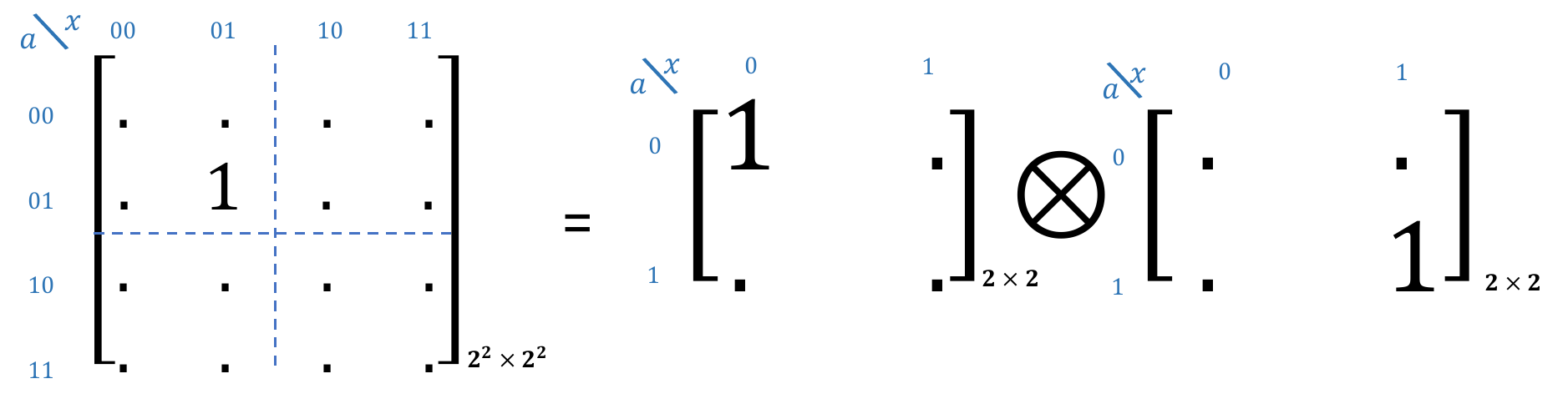}
	\caption{an example of decomposing a joint operation in a bipartite system. Given that $(01,01) \in \mathds{1}(\boldsymbol{D}_{\mathcal{CP}})$ and other elements are unknown. By applying the rules of the tensor product, we can deduce that $(0,0) \in \mathds{1}(\boldsymbol{D}_{\mathcal{CP}}^{(t_1)})$ and $(1,1) \in \mathds{1}(\boldsymbol{D}_{\mathcal{CP}}^{(t_2)})$. Thus, we have $\boldsymbol{D}_{\mathcal{CP}}=\boldsymbol{D}_{\mathcal{CP}}^{(t_1)} \otimes \boldsymbol{D}_{\mathcal{CP}}^{(t_2)}$, where $t_1 \in \{0,1\}$ and  $ t_2 \in \{1,3\}$.}
	\label{fig:block}
\end{figure}

 By the rules of  tensor product,  given $D_{\mathcal{CP}} ( A_1A_2\dots A_n| X_1X_2\dots X_n)=1$,  it follows that there exist $D_{\mathcal{CP}}^{(t_j)}(A_j|X_j)=1 $, for each $j\in \{1,\dots,n\}$ and $ t_j \in \{0,1,2,3\}$, such that Eq. (\ref{Appendeq:D}) holds for each possible index $ (\vec{a},\vec{x}) \in \mathds{1} \left( \boldsymbol{Z} \right)$. This relationship is illustrated in Fig.  \ref{fig:block}.

In Case 2, for any two distinct indices $(\vec{a}, \vec{x})$ and $(\vec{a}', \vec{x}')$ in $\mathds{1} (\boldsymbol{Z})$, there exists at least one party $S_q$, such that $a_q \neq a'_q$ but $x_q = x'_q$. Since a local operation is represented by a left stochastic matrix which is expressed in Eq. (\ref{Appendeq:singleD}), this implies that $(a_q, x_q)$ and $(a'_q, x'_q)$ cannot coexist within the same local operation. Consequently, we can derive Eq. (\ref{Appendeq:empty}).
 \begin{equation}
    \begin{split}
        &\{\boldsymbol{D}_{\mathcal{CP}}|(\vec{a},\vec{x}) \in \mathds{1} \left( \boldsymbol{D}_{\mathcal{CP}}  \right) \} 
        \\
		&= \{   \otimes_{j=1}^{n} \boldsymbol{D}_{\mathcal{CP}}^{(t_j)}| (a_{j},x_{j}) \in \mathds{1}(\boldsymbol{D}_{\mathcal{CP}}^{(t_j)}),   t_j  \in \{0,1,2,3\} \}	 
  \label{Appendeq:D}
    \end{split}
\end{equation}
\begin{equation}
     \begin{split}
         \sum_{a} D_{\mathcal{CP}}^{(t_q)} (a,x) &= D_{\mathcal{CP}}^{(t_q)} (0,x) +D_{\mathcal{CP}}^{(t_q)} (1,x) =1
         \\
         \forall x\in \{0,1\} & \text{ and } \forall t_q \in \{0,1,2,3\}
 	\label{Appendeq:singleD}
     \end{split}
 \end{equation}

Note that given a single index $(a_j, x_j)$,  where $j \in \{1, \dots, n\}$,  there are two local operations such that $ \boldsymbol{D}_{\mathcal{CP}}^{(t_{j})}(a_j, x_j)=1$ ($t_{j} \in \{0,1,2,3\}$). By applying Eqs. (\ref{Appendeq:empty})–(\ref{Appendeq:D}), we deduce that there are $|\mathds{1}(\boldsymbol{Z})|\times2^n$ deterministic operations which satisfy $|\mathds{1}(\boldsymbol{Z}) \cap \mathds{1}(\boldsymbol{D}_{\mathcal{CP}})| = 1$, and others satisfy $|\mathds{1}(\boldsymbol{Z}) \cap \mathds{1}(\boldsymbol{D}_{\mathcal{CP}})| = 0$. This completes the proof.

Considering the binary inputs and outputs, it can be deduced that if a matrix $\boldsymbol{Z}$ belongs to the set of extra effects $ E_{\mathcal{CP}}^{\mathcal{C}}$,  we have

\begin{cory}\!\!\textbf{.}\label{cory1}
     If a $2^n \times 2^n$ $\{0,1\}$-valued matrix $\boldsymbol{Z}$ with $ 0 \leq |\mathds{1} (\boldsymbol{Z})| < 2^n $ is necessarily an extra effect, it should meet the following condition.

    Case 3. when the number of nonzero elements satisfies $|\mathds{1} (\boldsymbol{Z})| \geq 2$, there exist at least two different indices $\left( \vec{a},\vec{x} \right),\left( \vec{a}',\vec{x}' \right) \in \mathds{1}(\boldsymbol{Z})$, where $S_{id}(\vec{x},\vec{x}') \subset \left\lbrace 1,\dots, n \right\rbrace  ( 0 \leq |S_{id}(\vec{x},\vec{x}')|   < n ) $, such that for any index $ q \in S_{id}(\vec{x},\vec{x}')$, we have ${a}_q = {a'}_q$.
\end{cory}

While Case 3 establishes a necessary condition for identifying an extra effect, we will also show that it serves as a sufficient condition for any extra effect.

\begin{clm}\!\!\textbf{.}\label{clm2}
    Any extra effect $\boldsymbol{e}_{\mathcal{CP}}^{\mathcal{C}}$ is invalid.
\end{clm}

It has been shown that any normal effect $\boldsymbol{e}_{\mathcal{CP}}$ is guaranteed to be a valid extremal effect~\cite{eftaxiasMultisystemMeasurementsGeneralized2023}. We now further conclude that these effects exhaust  valid extremal effects. Any ${0,1}$-valued effect outside the set $E_{\mathcal{CP}}$ is invalid.

Considering two indices $\left(\vec{a}, \vec{x}\right)$ and $(\vec{a}', \vec{x}') \in \mathds{1}(\boldsymbol{Z})$ in Corollary \ref{cory1}, there are two possible situations for each party $S_j\left( j \in \{1, \dots,n\} \right) $. Firstly, when $j \in S_{id}(\vec{x},\vec{x}')$, we have ${x}_j={x'}_j$ and ${a}_j={a'}_j$. It corresponds to two possible local operations, as said in the proof of Claim \ref{clm1}. Secondly, $j \notin S_{id}(\vec{x},\vec{x}')$ implies ${x}_j \ne x'_j$. Based on the fact that $ \boldsymbol{D}_{\mathcal{CP}}^{(t_j)}:\mathcal{X} \to \mathcal{A} ,t_j\in \{0,1,2,3\}$, we always find a local operation $\boldsymbol{D}_{\mathcal{CP}}^{(t_j)} $ that satisfy $( {a}_j,{x}_j) \in \mathds{1}( \boldsymbol{D}_{\mathcal{CP}}^{(t_j)})	 \text{ and } ( {a'}_j,{x'}_j) \in \mathds{1}( \boldsymbol{D}_{\mathcal{CP}}^{(t_j)})$. 

Similar to the proof in Claim \ref{clm1}, we use Eq. (\ref{Appendeq:D}) to infer that for any $\boldsymbol{Z}$ satisfying Corollary \ref{cory1}, there exists at least one local operation $\boldsymbol{D}_{\mathcal{CP}}$, such that $\left( \vec{a},\vec{x}\right) \in \mathds{1}( \boldsymbol{D}_{\mathcal{CP}})	 \text{ and } \left( \vec{a}', \vec{x}'\right) \in \mathds{1}( \boldsymbol{D}_{\mathcal{CP}})$. Therefore, by invoking the inner product of two matrices, we have

\begin{equation}
	\boldsymbol{Z} \cdot \boldsymbol{D}_{\mathcal{CP}}  \geq 2, \exists \boldsymbol{D}_{\mathcal{CP}}
	\label{Appendeq:Ec}
\end{equation}

According to Eq. (\ref{eq:E_nor}), $\boldsymbol{Z}$ must belong to $E_{\mathcal{CP}}^{\mathcal{C}}$. Therefore, a $\{0,1\}$-valued matrix $\boldsymbol{Z}$ is a valid extremal effect if and only if it satisfies one of the conditions specified in Claim \ref{clm1}.

\begin{thm}\!\!\textbf{.}\label{thm3}
     In both box world and local theory, a spanning measurement vector is valid if and only if it is $\{0,1\}$-valued.
\end{thm}

In Claim \ref{clm2}, we demonstrate that any matrix $\boldsymbol{Z}$ in Case 3 represents an extra effect, $\boldsymbol{e}_{\mathcal{CP}}^{\mathcal{C}}$, which is not physically valid in the box world and local theory. This implies that probabilistic classical processes need carefully adjusted weights, since even small changes can lead to invalid measurements. As highlighted by previous work~\cite{woodLessonCausalDiscovery2015a}, such processes lack the stability needed for reliable realization,  thereby limiting their applicability in practical scenarios.

Theorem \ref{thm3} restricts our discussion to the set of deterministic classical processes $\{\boldsymbol{M}_{\mathcal{CP}}^{D}\}$. This subset encompasses processes where the logical consistent causality of the observed variables remains robust even when causal parameters are modified.

\section{Measurements and dynamical causal order with four parties}

By focusing on the deterministic set, we avoid the implausible fine-tuning that arise in extremal probabilistic processes, provides a logically consistent way for analyzing the behavior of measurements in multipartite systems. However, as the system size increases, describing the causal order in a classical process becomes increasingly challenging due to the exponential growth of Boolean functions, it is difficult to obtain all details about each possible classical process. To learn classical processes in a multipartite scenario, we approach them from multiple levels. 

In classical processes, causal information is typically represented by a directed graph that outlines causal pasts and futures, known as the causal structure~\cite{tselentisAdmissibleCausalStructures2023a}. Due to the variety of Boolean functions, multiple valid Boolean functions can correspond to  the same causal structure. By applying different value of $\vec{a}$ to each Boolean function, the classical process simulates various possible  $\vec{x}$ consistent with the causal order, generating a complete truth table. At this stage, certain information is reduced by symmetries, such as the relabeling of parties, inputs, and outputs. Finally, by constructing the symmetry group, we identify all corresponding vertices, obtaining more information than just the causal structure. Using this method, we analyze the set of deterministic classical processes in the $(4,2,2)$ scenario.

Because the vertex enumeration problem for a polytope is NP-hard, making it impractical to obtain the complete set of vertices in a 4-partite system using solvers such as \textsc{panda}~\cite{PANDA}. To compute the vertices of the polytope representing deterministic classical processes in the $(4,2,2)$ scenario, we utilize integer linear programming (ILP) techniques. By optimizing random objective functions approximately two billion times, we identified $5,541,744$ integer vertices of the polytope characterizing classical processes. After removing symmetry, these vertices can be grouped into $1,291$ distinct classes. We also confirmed that these $1,291$ equivalent classes produce $5,541,744$ vertices without any additional points, as verified by generating the symmetry group. Therefore, we have identified all vertices within the $1,291$ sets.
\begin{figure}[htbp]
    \centering
    \includegraphics[scale=0.4]{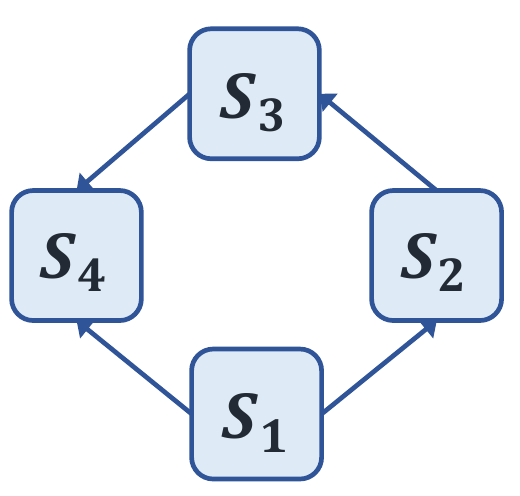}
    \caption{A fixed causal order in a 4-partite system. $S_1$ represents the global causal past, receiving a constant from the environment. There is a causal order chain among $S_1$, $S_2$, $S_3$ and $S_4$, where each  $S_i$ s in the causal past of $S_{i+1}$ for $i \in  \{1,2,3\}$.  Additionally, $S_4$ is jointly influenced by $S_1$ and $S_3$.}
    \label{fig:party4_a1}
\end{figure}
\begin{figure}[htbp]
    \centering
    \includegraphics[scale=0.33]{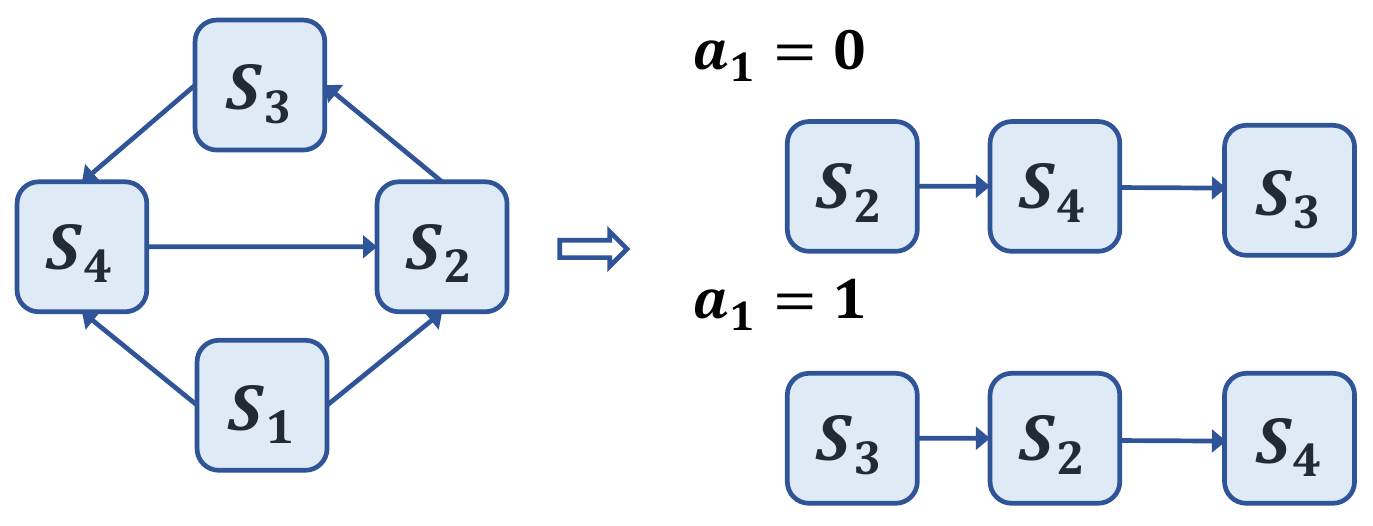}
    \caption{An adaptive causal order in a 4-partite system. $S_1$ dynamically influences the causal orders between $S_2$, $S_3$ and $S_4$. If the output of $S_1$ is $0$, the upper right channel chain is selected. Conversely, if the output is $1$, the lower right channel chain is chosen.}
    \label{fig:party4_a2}
\end{figure}
\begin{figure*}[]
    \includegraphics[scale=0.45]{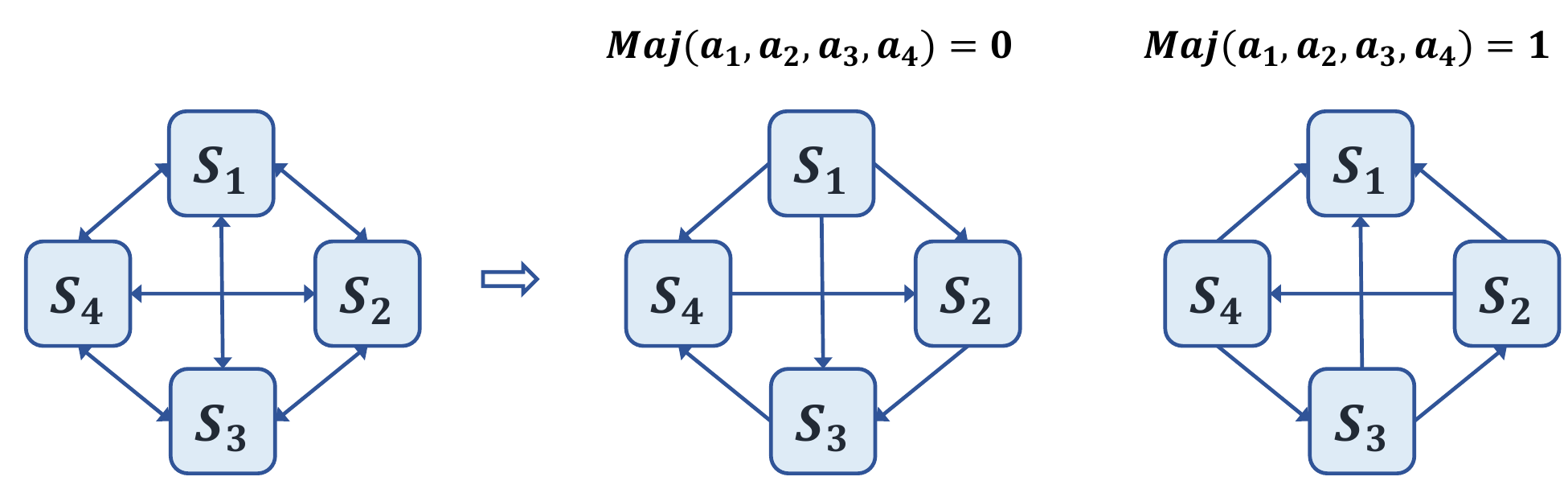}
    \caption{one of indefinite causal orders is represented by a complete graph. Each double arrow is divided into two cases based on the majority function. When the number of $0$s of outputs is greater than or equal to $2$, we have $\text{Maj}(a_1, a_2, a_3, a_4)=0$. If $\text{Maj}(a_1, a_2, a_3, a_4)=0$, the causal structure is $x_1=1$, $x_2=\bar{a_1} \vee \bar{a_3}$, $x_3=\bar{a_1} \vee \bar{a_4}$, and $x_4=\bar{a_1} \vee \bar{a_2}$. Otherwise, $x_1=\bar{a_2} \vee \bar{a_3} \vee \bar{a_4}$, $x_2=a_4$, $x_3=a_2$, and $x_4=a_3$.
}
\label{fig:party4_com}
\end{figure*}

By analyzing the causal order from these vertices, we find that if party $S_i$ causally effects party $S_j$, the output of $S_i$ will be correlated with the input of $S_j$. Consequently, if we flip the output of $S_i$ while keeping the outputs of the other parties fixed, the input of party $S_j$ would also be flipped. By enumerating the outputs of each party, we have obtained a causal structure for each class. Note that different Boolean functions can result in a isomorphic causal structure,thus $1,291$  classes yield to $69$ non-isomorphic classes. 

Furthermore, we show that these $69$ classes of causal structures encompass all Siblings-on-Cycles (SOCs) graphs in 4-partite systems ~\cite{tselentisAdmissibleCausalStructures2023a}. This finding confirms the conjecture in ~\cite{tselentisAdmissibleCausalStructures2023a}, where the set of SOC graphs in the $(4,2,2)$ case is sufficiently valid. Importantly, the necessary criterion of the SOC graphs ensures that the collection of $5,541,744$ vertices comprises all possible causal structures. This guarantees that we have successfully obtained all the vertices within the polytope.  In order to systematically study the causal order in a multipartite classical system, the $69$ classes of causal structures can be roughly classified into 3 types. 

The first type consists of fixed causal orders, where the sequencing of events remains consistent in any possible Boolean function~\cite{oreshkovCausalCausallySeparable2016}. There exists necessarily at least one party that receives a constant from the environment in such a causal structure. That is, the global past is predefined. In Fig.  \ref{fig:party4_a1}, $2$ out of $1,291$ classes yield this isomorphic causal structure, both of which have fixed causal orders. For example, one of the vertices is represented as
\begin{equation}
    x_1=1, \quad x_2=\bar{a}_1, \quad x_3=\bar{a}_2, \quad x_4=a_1 \oplus a_3
    \label{eq:party4_a1}
\end{equation}

Secondly, 4-partite systems can develop more adaptive causal orders. It means that the directions of causal order between certain parties change based on the outputs of others. We know that the presence of a unidirectional cycle leads to the causal paradox in a tripartite classical system~\cite{baumelerSpaceLogicallyConsistent2016}. However, by introducing a global past, for a 4-partite system, some of adaptive causal orders that arise can help resolve the inconsistencies that may occur in the tripartite system~\cite{tselentisAdmissibleCausalStructures2023a}. For example, in Fig.  \ref{fig:party4_a2}, there is a directed cycle between $S_2$, $S_3$ and $S_4$, which individually leads to a causal paradox in a tripartite system. Given $S_1$ as a control party, the causal loop is broken into a mixture of two fixed causal orders, i.e., $S_2 \preceq S_4 \preceq S_3$ and $S_3 \preceq S_2 \preceq S_4$. The behavior of an example is described as Eq. (\ref{eq:party4_a2}). 
\begin{equation}
    x_1=1, \quad x_2=\Bar{a}_1\vee \Bar{a}_3, \quad x_3=a_1\vee \Bar{a}_4, \quad x_4=\Bar{a}_2
    \label{eq:party4_a2}
\end{equation}

The third type is a set of indefinite causal orders (ICOs), which signifies that the causal structure is not predefined. In this framework, each party receives a variable $x$ from the environment and behaves consistently based on that information. ICOs are a crucial concept in quantum information processing and quantum communication, as they challenge our classical understanding of causality. Note that there are $15$ out of the $69$ causal structures that show ICOs, which is higher than $1$ out of $7$ in a tripartite system. This indicates that the advantages demonstrated by ICOs will become increasingly normal in multipartite systems. The typical one is a completely connected graph. Similar to the self-circle in $(3,2,2)$ scenario,  one of the simulations is divided into two causal orders depending on the majority of outputs of all parties. For example, one behavior is expressed by Eq. (\ref{eq:party4_com}), and shown in Fig.  \ref{fig:party4_com}.
\begin{equation}
    \begin{split}
   &x_1=\bar{a}_2 \vee \bar{a}_3 \vee \bar{a}_4, \quad x_2=\bar{a}_1\vee \bar{a}_3 \vee a_4, 
   \\
   &x_3=\bar{a}_1\vee \bar{a}_4 \vee a_2, \quad x_4=\bar{a}_1\vee \bar{a}_2 \vee a_3
    \label{eq:party4_com}        
    \end{split}
\end{equation}

\section{Dynamical causal order and its device-independent certification}
\subsection{A 4-partite parallel-serial switch}
\begin{figure}
    \centering
    \includegraphics[scale=0.42]{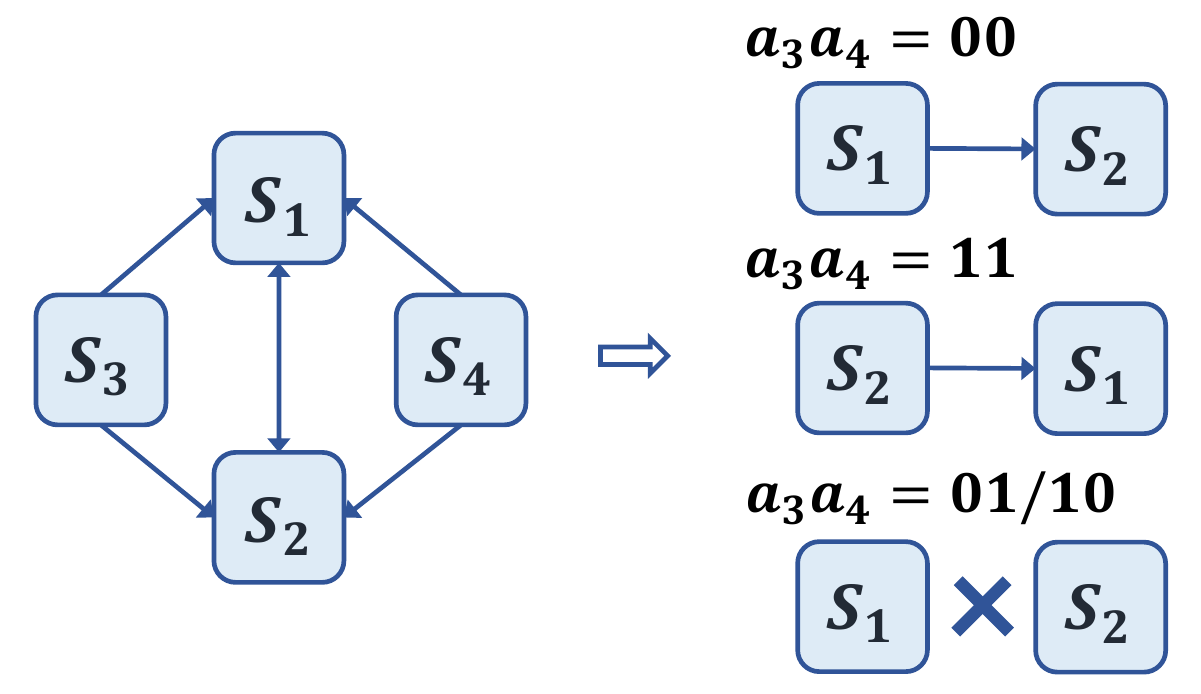}
    \caption{Causal structure of PAR-SER switch. The one-way signaling causal order $S_1 \preceq S_2(S_2 
 \preceq S_1$) is allowed when the outputs of control parties $S_3,S_4$ are $00(11)$. In other instances, $S_1,S_2$ are no-signaling. }
    \label{fig:switch}
\end{figure}

Fig.  ~\ref{fig:switch} depicts one of the $69$ classes of causal structures in 4-partite classical processes, which can be seen as an extension of the quantum switch on classical systems, enabling the configuration of signaling channels either 
independently or sequentially. 

Such causal structures, incorporating causal nonseparability~\cite{AraujoWitnessingcausalnonseparability2015,feixCausallyNonseparableProcesses2016}, has recently moved beyond theoretical curiosity and has been shown to confer experimental advantages in certain tasks~\cite{RennerComputationalAdvantageQuantum2022,Unitarychanneldiscrimination,BavarescoStrictHierarchyParallel2021,ZhaoQuantumMetrologyIndefinite2020a,EblerEnhancedCommunicationAssistance2018,GuerinExponentialCommunicationComplexity2016,FeixQuantumsuperpositionorder2015,AraujoComputationalAdvantageQuantumControlled2014b,ChiribellaPerfectdiscriminationnosignalling2012b,ChapeauNoisyquantumindefinitecausalorder2021,YinExperimentalsuperHeisenbergquantum2023,baumelerFlowDynamicalCausal2024}. However, before experiments can be devised, the desired causally non-separable process needs to be certified. This can be achieved by violating a causal inequality, analogous to the demonstration of nonlocality by violating Bell inequalities~\cite{oreshkovQuantumCorrelationsNo2012,Branciardsimplestcausalinequalities2015b, baumelerSpaceLogicallyConsistent2016,BaumelerPerfectsignalingthree2014d,baumelerMaximalIncompatibilityLocally2014}. Alas, not all causally non-separable processes can be confirmed in a device-independent manner~\cite{AraujoWitnessingcausalnonseparability2015,oreshkovCausalCausallySeparable2016}. One prominent counterexample is the quantum switch ~\cite{ChiribellaQuantumcomputationsdefinite2013c}.  Recently, van der Lugt et al. introduced a device-independent method to certify the indefinite causal orders in the quantum switch ~\cite{VanDerLugtDeviceindependentcertificationindefinite2023a}. Their method forms the basis of our approach to device-independent certification of causal nonseparability.

In Fig.  ~\ref{fig:switch}, which from now on we call the parallel-serial switch (PAR-SER switch), control over signaling channels between two parties in either a parallel or serial configuration is achieved by two of the parties. Specifically, a one-way signaling channel from $ S_1$  to  $S_2$  ( $S_1 \preceq S_2$ ) arises when control parties  output  $a_3a_4=00$ . Conversely, the channel reverses to  $S_2 \preceq S_1$  when $a_3a_4 = 11 $. In other scenarios ($ a_3 \oplus a_4 = 1 $), signaling is prohibited ( $S_1 \npreceq \nsucceq S_2$ ). The behavior of  PAR-SER switch is shown in Eq. (\ref{Aqqdeq:switch})
\begin{equation}
        x_1=a_2 \vee \Bar{a}_3 \vee \Bar{a}_4, \quad x_2=a_1 \vee a_3 \vee a_4, \quad x_3=0, \quad x_4=0 \label{Aqqdeq:switch}
\end{equation}

Within the process matrix framework, the PAR-SER switch in a classical system is represented as a positive diagonal matrix, utilizing the identity $\mathbb{I}$ and the Pauli matrix $\sigma_{z}$~\cite{baumelerMaximalIncompatibilityLocally2014,baumelerSpaceLogicallyConsistent2016}. The matrix $W_{PAR-SER}$ is written as Eq. (\ref{Appdeq:par-ser}). Here, each term in Eq. (\ref{Appdeq:par-ser}) is a tensor product of input space and output spaces. The identity $\mathbb{I}$ represents $\mathbb{I}^{\otimes 4}$, $\sigma^{m_i}$ indicates a Pauli matrix $\sigma_z$ applied to the $i$-th input ($m=x$) space or output ($m=a$) space. For example, $\sigma^{x_1,x_3} \otimes \sigma^{a_2} = \sigma_{z}\otimes \mathbb{I} \otimes \sigma_{z}\otimes \mathbb{I} \otimes \mathbb{I}\otimes \sigma_{z}\otimes \mathbb{I}\otimes \mathbb{I}$.
\begin{widetext}
\begin{equation}
\begin{split}
    &W_{PAR-SER}= \frac{1}{2^4} [ \mathbb{I}\otimes \mathbb{I}+ \sigma^{x_3}  \otimes \mathbb{I}+ \sigma^{x_4} \otimes \mathbb{I}+ \sigma^{x_3,x_4} \otimes\mathbb{I} +\frac{1}{4}( \sigma^{x_1}+\sigma^{x_1,x_3}+ \sigma^{x_1,x_4}+\sigma^{x_1,x_3,x_4} ) \otimes
    \\
   & (-3 \mathbb{I} +\sigma^{a_2}-\sigma^{a_3}-\sigma^{a_4} 
   -\sigma^{a_2,a_3}-\sigma^{a_2,a_4}+\sigma^{a_3,a_4}+\sigma^{a_2,a_3,a_4}) +\frac{1}{4}  ( \sigma^{x_2}+\sigma^{x_2,x_3}+ \sigma^{x_2,x_4}+\sigma^{x_2,x_3,x_4} ) \otimes
    \\
     & (-3 \mathbb{I} +\sigma^{a_1}+\sigma^{a_3}+\sigma^{a_4}+\sigma^{a_1,a_3}
     +\sigma^{a_1,a_4}+\sigma^{a_3,a_4}+\sigma^{a_1,a_3,a_4}) + \frac{1}{4}  ( \sigma^{x_1,x_2}+\sigma^{x_1,x_2,x_3}+ \sigma^{x_1,x_2,x_4}+\sigma^{x_1,x_2,x_3,x_4} )
  \otimes
     \\
   &  (2 \mathbb{I}- \sigma^{a_1}- \sigma^{a_2}-2\sigma^{a_3,a_4}+\sigma^{a_2,a_3}
     +\sigma^{a_2,a_4}-\sigma^{a_1,a_3}-\sigma^{a_1,a_4}-\sigma^{a_1,a_3,a_4}-\sigma^{a_2,a_3,a_4}) ]
     \label{Appdeq:par-ser}
\end{split}
\end{equation}
\end{widetext}

\subsection{Divice-independent certification of the PAR-SER switch}
\begin{figure*}
    \centering
    \includegraphics[scale=0.2]{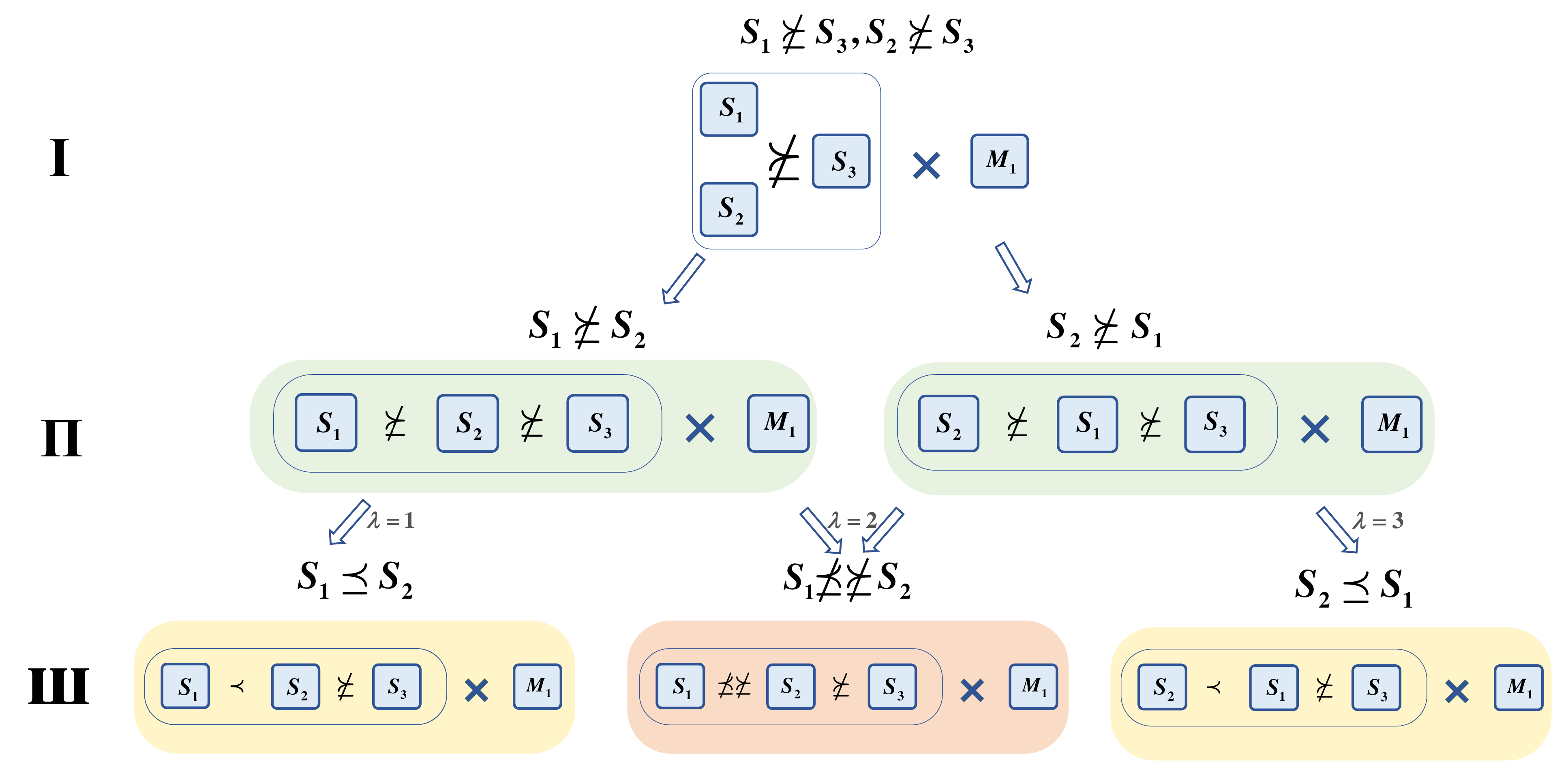}
    \caption{Based on experimental settings, correlations can be divided into three levels by the causal order classification criteria. \uppercase\expandafter{\romannumeral1}. The basic setting. The players $S_1$, $S_2$, and $S_3$ interact with an auxiliary system $M_1$. Here, $S_3$ is in the causal future of $S_1$ and $S_2$, denoted by $S_1\nsucceq S_3$ and $S_2\nsucceq S_3$. Additionally, $M_1$ is spacelike-separated from the three players. \uppercase\expandafter{\romannumeral2}. The relativistic causality between $S_1$ and $S_2$ (green block). When unidirectional signals are sent from $S_1$ to $S_2$, $S_1$ is not affected by $S_2$, as $S_2$ can only influence events within its own forward light cone. This relationship is denoted by $S_1 \nsucceq S_2$. Similarly, the reverse holds for $S_2$, expressed as $S_2 \nsucceq S_1$. \uppercase\expandafter{\romannumeral3}. The definite causal orders. In each run of experiment, three definite causal orders are probabilistically determined by $\lambda$. If signaling is allowed, the channel from $S_1$ to $S_2$ corresponds to $S_1 \preceq S_2$, and from $S_2$ to $S_1$ corresponds to $S_2 \preceq S_1$ (yellow block). In a no-signaling scenario, this relationship is represented as $S_1 \npreceq \nsucceq S_2$ (orange block).}
    \label{fig:causal_tree}
\end{figure*}

For simplicity in the analysis, the two control parties,  $S_3$ and $S_4$, are replaced by a single party, $S_3$, which operates with three different control states. To identify the adaptive causal order in PAR-SER switch, we design an experiment involving four parties: $S_1$, $S_2$, $S_3$, and $M_1$, where $M_1$ serves as an auxiliary system correlated with the control system $S_3$ (see Fig.  \ref{fig:causal_tree}). The trinary input and output of the auxiliary system $M_1$ are denoted by $y$ and $b$, respectively.

The experiment is designed as follows. interventions by  $S_3$  occur within the future light cone of those by  $S_1$ and  $S_2$, while  $M_1$  remains spacelike-separated from the other parties (see part \uppercase\expandafter{\romannumeral1} in Fig.  \ref{fig:causal_tree}). Similar to the set of local hidden variable correlations, these correlations are constrained by three assumptions: definite causal order, relativistic causality, and free interventions ~\cite{VanDerLugtDeviceindependentcertificationindefinite2023a}. The first assumption posits that the correlation between $S_1$ and $S_2$ probabilistically establishes a definite causal order in each run of the experiment, depending on the value of a variable $\lambda$. Secondly, relativistic causality  imposes two one-way signaling constraints between $S_1$ and $S_2$:  $S_1$($S_2$) could send her input and output to $S_2$($S_1$), but not vice versa. The corresponding set of correlations is denoted by $\mathcal{P}^{S_1 \nsucceq S_2}$ ($\mathcal{P}^{S_2 \nsucceq S_1}$) (see part \uppercase\expandafter{\romannumeral2} in Fig.  \ref{fig:causal_tree}). These can be further classified in the experiments. In each run, $S_1$ and $S_2$ are compatible with one of three definite causal orders: $S_1 \preceq S_2$, $S_2 \preceq S_1$, or $S_1 \npreceq\nsucceq S_2$ (see part \uppercase\expandafter{\romannumeral3} in Fig.  \ref{fig:causal_tree}). The corresponding sets of correlations are denoted by $\mathcal{P}^{S_1 \preceq S_2}$, $\mathcal{P}^{S_2 \preceq S_1}$, and $\mathcal{P}^{S_1 \npreceq\nsucceq S_2}$, respectively. Finally, note that the hidden variable is statistically independent, and parties can not send signal outside their future light cones. This condition imposes additional constraints on the conditional probability distribution within the system. 

The sets of correlations generated in this experimental setting are expressed as Eqs. (\ref{Appdeq:p(ab)})-(\ref{Appdeq:drf2}), corresponding to part \uppercase\expandafter{\romannumeral1} and \uppercase\expandafter{\romannumeral2} in Fig.  \ref{fig:causal_tree}. Note that the set of no-signaling correlations $\mathcal{P}^{S_1 \npreceq \nsucceq S_2}$ is compatible with both one-way signaling processes. Here $\mathcal{P}_{\Vec{a}b|\Vec{x}y}$ is a set of positive and normalized correlations, and $\bot_{p}$ represents statistical independence. For example, $\Vec{a}\bot_{p} y$ means that for all $\Vec{a},b,\Vec{x},y,y'(y'\ne y)$, we have $\sum_{b}p(\Vec{a}b|\Vec{x}y) = \sum_{b}p(\Vec{a}b|\Vec{x}y')$. 

\begin{align}
    \begin{split}
        p(\Vec{a}b|\Vec{x}y)&=\sum_{\lambda \in \{1,2,3\} } p(\lambda)p(\Vec{a}b|\Vec{x}y \lambda)
        \\
        p(\lambda) \geq 0, &\sum_{\lambda\in\{0,1\}} p(\lambda)=1 , \forall p(\Vec{a}b|\Vec{x}y)\in \mathcal{P}_{\Vec{a}b|\Vec{x}y}  
            \label{Appdeq:p(ab)}
    \end{split}
    \\
        \mathcal{P}:&=\{p \in \mathcal{P}_{\Vec{a}b|\Vec{x}y}: \Vec{a}\bot_{p} y,  b \bot_{p} \Vec{x} \}
            \label{Appdeq:ns}
         \\
        \mathcal{P}^{S_1 \nsucceq S_2}:&=\{ p \in \mathcal{P}: a_1b \bot_{p} x_2, \{a_1,a_2,b\} \bot_{p} x_3\}
            \label{Appdeq:drf1}
        \\
        \mathcal{P}^{S_2 \nsucceq S_1}:&=\{ p \in \mathcal{P}: a_2b \bot_{p} x_1, \{a_1,a_2,b\}\bot_{p} x_3\}
            \label{Appdeq:drf2}
\end{align}

By this experimental setup, the set of possible correlations, denoted by $\mathcal{CO}$, can be written as a convex mixture of two one-way signaling causal orders. It means that $\mathcal{CO}$ is causal separable (see Eq. (\ref{Appdeq:drf})).
\begin{equation}
    \begin{split}
        &\mathcal{CO} =q_1\mathcal{P}^{S_1 \preceq S_2} + q_2 \mathcal{P}^{S_2 \preceq S_1}+  q_3 \mathcal{P}^{S_1 \npreceq \nsucceq S_2}
        \\
        &= p\mathcal{P}^{S_1 \nsucceq S_2} + (1-p) \mathcal{P}^{S_2 \nsucceq S_1}, \quad
        \sum_{i=1}^{3}q_i=1, q_i\geq 0, p \geq 0 
    \label{Appdeq:drf}
    \end{split}
\end{equation}

\begin{thm}\!\!\textbf{.}\label{thm4}
      The causal correlations  $\mathcal{CO}$ satisfy 
\begin{equation}
    \begin{split}         
        &p(b=0,x_2(a_2\oplus x_1)=0|y=0)+   
        \\
        &p(b=1,x_1(a_1\oplus x_2)=0|y=0)+  
        \\
        &p(b=2,F(a_1,a_2|x_1,x_2)=1|y=0)+  
        \\
        &\frac{1}{4(3+\sqrt{3})}I_3(a_3,b|x_1=x_2=0,x_3,y) 
        \leq \frac{7}{8}+\frac{1}{2\sqrt{3}}
    \label{Appdeq:inequa}
    \end{split}
\end{equation}
\end{thm}
Here, $F(a_1,a_2|x_1,x_2)$ is a guess game where there are two cooperating players $S_1,S_2$ (see Eq. (\ref{Appdeq:F}), where $\odot$ is Xnor and $\oplus$ is Xor). $I_3(a_3,b|x_1=x_2=0,x_3,y)$ is a tailored CGLMP inequalities based on the condition $x_1=x_2=0$. It is maximally violated by maximally entangled state in a qutrit system (see Eq. (\ref{Appdeq:I3}))~\cite{SalavrakosBellInequalitiesTailored2017}. 

In the guess game $F(a_1,a_2|x_1,x_2)$, every player receives one specific question which depends on the inputs of $S_1$ and $S_2$. If $x_1\odot x_2=1$, two players are supposed to give the same answer, i.e., $a_1=a_2$. Nevertheless, they have to guess each other's input when $x_1\oplus x_2=1$. $F(a_1,a_2|x_1,x_2)$ is a signaling causal inequality where the bounds for signaling correlatrions, $\mathcal{P}^{S_1\preceq S_2}$ and $\mathcal{P}^{S_2\preceq S_1}$, are given by $\frac{3}{4}$. While, the players can perfectly win in no-signling scenario.

\begin{equation}
    F= (x_1 \oplus x_2)  (a_1 \odot x_2)  (a_2 \odot x_1 )+(x_1 \odot x_2) ( a_1\odot a_2)
    \label{Appdeq:F}
\end{equation}
\begin{widetext}
    \begin{equation}
    \begin{split}
        &I_3=\frac{1}{\sqrt{3}}[p(a_3=b|x_1=x_2=0, x_3=0,y=0)+p(a_3=b|x_1=x_2=0, x_3=1,y=0)+p(a_3=b|x_1=x_2=0, x_3=1,y=1)
        \\
        &+p(a_3+1=b|x_1=x_2=0, x_3=0,y=1)]+\frac{3-\sqrt{3}}{6}[p(a_3=b-1|x_1=x_2=0, x_3=0,y=0)
        \\
        &+p(a_3=b-1|x_1=x_2=0, x_3=1,y=1)+p(a_3=b|x_1=x_2=0, x_3=0,y=1)+p(a_3-1=b|x_1=x_2=0, x_3=1,y=0)]
        \label{Appdeq:I3}
    \end{split}
\end{equation}
\end{widetext}

Note that each term in Eq. (\ref{Appdeq:inequa}) certifies a specific feature in  PAR-SER switch. The first two terms are LGYNI game~\cite{Branciardsimplestcausalinequalities2015b}, which identity two signaling causal orders with opposite directions. The third term certifies the no-signaling condition. And the fourth term detects nonlocality between $S_3$ and $M_1$. In the prepare-and-measure scenarios, when conditioned on $x_1=x_2=0$, the target particles reach the same final state in each run of the experiment. Therefore, the three causal orders shown in Fig.  \ref{fig:causal_tree} become indistinguishable~\cite{VanDerLugtDeviceindependentcertificationindefinite2023a}.  As a result, the correlation $p(\vec{a}b|\vec{x}y)$ reduced to a bipartite system, i.e., $p(\Vec{a}b|00x_3y)=p(a_3b|x_3y)$. To prove Theorem \ref{thm4}, we first focus on the last term.

\begin{clm}\!\!\textbf{.}\label{clm3}
    If $p(a_3b|x_3y)$ is no-signaling correlation, we have
\begin{equation}
    \frac{1}{4(3+\sqrt{3})}I_3(a_3,b|x_3,y) \leq \frac{1}{8}+\frac{1}{2\sqrt{3}} -\frac{1}{4}p(b=0|y=0)
    \label{Appdeq:I3ineq}
\end{equation}
\end{clm}

In general, the no signaling correlation is written as 
\begin{equation}
    p=\sum_{i}p_i p^{L}_{i} + \sum_{j}q_j p^{NS}_{j}
\end{equation}
where $\sum_{i}p_i+\sum_{j}q_j=1$ and $q_i,p_j\geq 0$ for all $i,j$. $p_i^{L}$ are local extremal correlations, and $p_j^{NS}$ are nonlocal extremal correlations, respectively. By the marginal probability distributions in a (2,2,3) scenario, one obtain
\begin{equation}
\begin{split}
        p(b=0|y=0) \leq 1 \cdot \beta+ \frac{1}{2}\cdot (1-\beta) 
        =\frac{1}{2}+\frac{1}{2}\beta
    \label{Appdeq:p(b)}
\end{split}
\end{equation}
where $\beta$ is the fraction of locality in any no-signaling correlation, i.e., $\beta:= \sum_{i}p_i$. By knowing the bounds for local correlations and no-signaling correlations in tailored CGLMP inequality~\cite{SalavrakosBellInequalitiesTailored2017}, the upper bound of $\frac{1}{4(3+\sqrt{3})}I_3(a_3,b|x_3,y)$ reads
\begin{equation}
\begin{split}
        &\frac{1}{4(3+\sqrt{3})}I_3(a_3,b|x_3,y) 
        \\
        &\leq \frac{1}{4(3+\sqrt{3})} [\frac{1+3\sqrt{3}}{2}\cdot \beta +(2+2\sqrt{3}) \cdot (1-\beta)]
        \\
        &= \frac{1}{2\sqrt{3}} - \frac{1}{8}\beta
        \\
        &\leq \frac{1}{8}+\frac{1}{2\sqrt{3}}-\frac{1}{4}p(b=0|y=0)
\end{split}
\end{equation}
The last inequality follows by combining Eq. (\ref{Appdeq:p(b)}).

In the following, we complete the proof of Theorem \ref{thm4}.
By utilizing the linearity of convex set and Eq. (\ref{Appdeq:drf}), we can express the bound of Eq. (\ref{Appdeq:inequa}) for each set of causal correlations individually.

Given a conditional probability $p \in \mathcal{P}^{S_1 \preceq S_2 }$, we have 
\begin{equation}
    p(b=0,x_2(a_2\oplus x_1)=0|y=0) \leq p(b=0|y=0) \label{Appdeq:b0}
\end{equation}
Considering LGYNI game and guess game in the case of $S_1 \preceq S_2$, we have
\begin{align}
    &p(b=1,x_2(a_1\oplus x_2)=0|y=0) = \frac{3}{4} p(b=1|y=0) \label{Appdeq:b1}
    \\
    &p(b=2,F(a_1,a_2|x_1,x_2)=1|y=0) \leq \frac{3}{4} p(b=2|y=0)\label{Appdeq:b2}
\end{align}

Suppose that the first three terms of the inequality Eq. (\ref{Appdeq:inequa}) is denoted by $\alpha$. 
\begin{equation}
    \begin{split}
        \alpha&=p(b=0,x_2(a_2\oplus x_1)=0|y=0)
        \\
        &+p(b=1,x_2(a_1\oplus x_2)=0|y=0)
        \\
        &+p(b=2,F(a_1,a_2|x_1,x_2)=1|y=0)
        \label{Appdeq:alphanota}
    \end{split}
\end{equation}

Combinig with Eqs. (\ref{Appdeq:b0})- (\ref{Appdeq:b2}), we have
\begin{equation}
    \alpha \leq \frac{3}{4}+\frac{1}{4}p(b=0|y=0)\label{Appdeq:alpha}
\end{equation}

Applying Claim \ref{clm3} to the correlation $p(a_3b|x_1=x_2=0,x_3,y)$, the proof is complete.
\begin{equation}
\begin{split}
    &\alpha +\frac{1}{4(3+\sqrt{3})}I_3(a_3,b|x_1=x_2=0,x_3,y)
    \\
    &\leq \frac{7}{8}+\frac{1}{2\sqrt{3}}
\end{split}
\end{equation}

Similarly, the bound also hold for the other two sets of causal correlations. The proof is complete.

The causal inequality in Eq. (\ref{Appdeq:inequa}) is designed to certify a causally non-separable process, which can coherently control three causal orders in the PAR-SER switch. To explore this, we construct a quantum version of the PAR-SER switch, assuming that each party locally satisfies quantum mechanics.

The quantum PAR-SER switch operates as follows. Consider two target particles, $T_1$ and $T_2$, with initial states $|\psi\rangle_{T_1}$ and $|\psi\rangle_{T_2}$, respectively. Control qutrit $S_3$ is prepared in a state $\frac{|0\rangle + |1\rangle+|2\rangle}{\sqrt{3}}$. The target particles are transformed by two sequential operations, where the order of operations depends on $S_3$. 

\begin{figure}[htbp] 
	\centering 
	\vspace{-0.35cm} 
	\subfigtopskip=2pt %subfug-up
	\subfigcapskip=-5pt %subfig-subtitle
	\subfigure[If the state of $S_3$ is $|0\rangle$, $T_1$ goes through one-way signaling channel $S_1 \preceq S_2$, and $T_2$ keeps unchanged. ]{
 
		\label{fig:Q_PARswitch.AB}
		\includegraphics[scale=0.2]{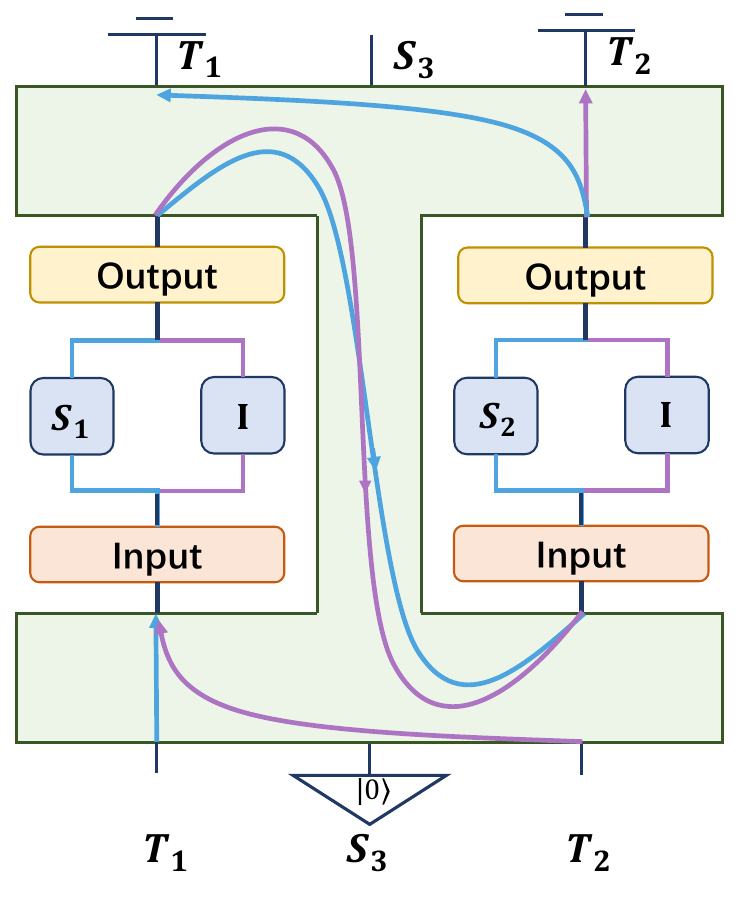}}
	\quad 
	\subfigure[If the state of $S_3$ is $|1\rangle$,  $T_1$ goes through one-way signaling channel $S_2 \preceq S_1$, and $T_2$ keeps unchanged.]{
		\label{fig:Q_PARswitch.BA}
		\includegraphics[scale=0.2]{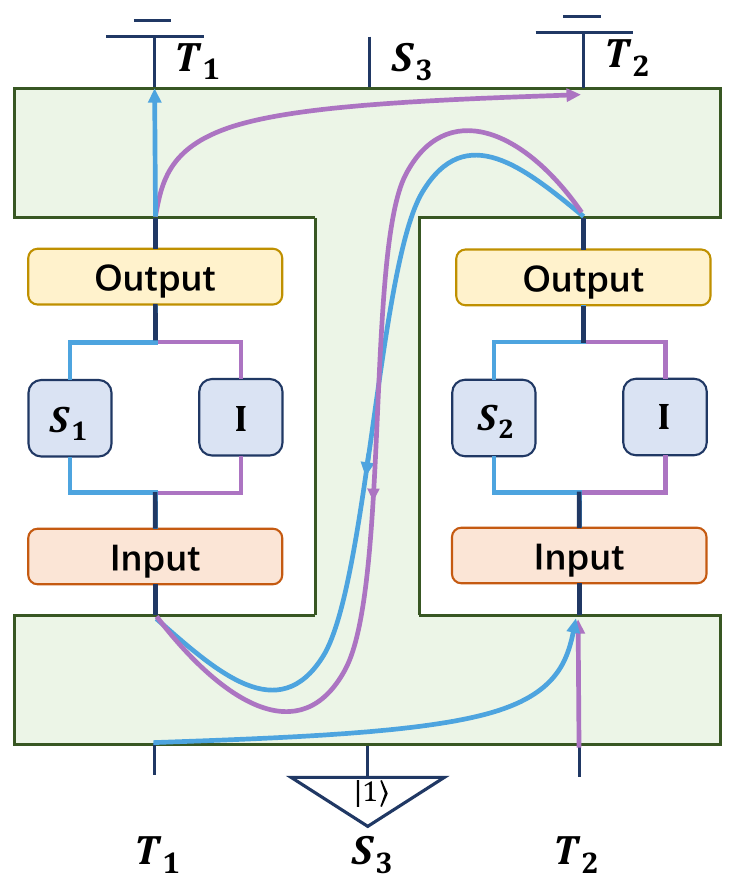}}
	\subfigure[If the state of $S_3$ is $|2\rangle$, the switch forbids signaling. $T_1$ and $T_2$ is transformed by the operation conducted by $S_1$ and $S_2$, respectively.]{
		\label{fig:Q_PARswitch.NS}
		\includegraphics[scale=0.2]{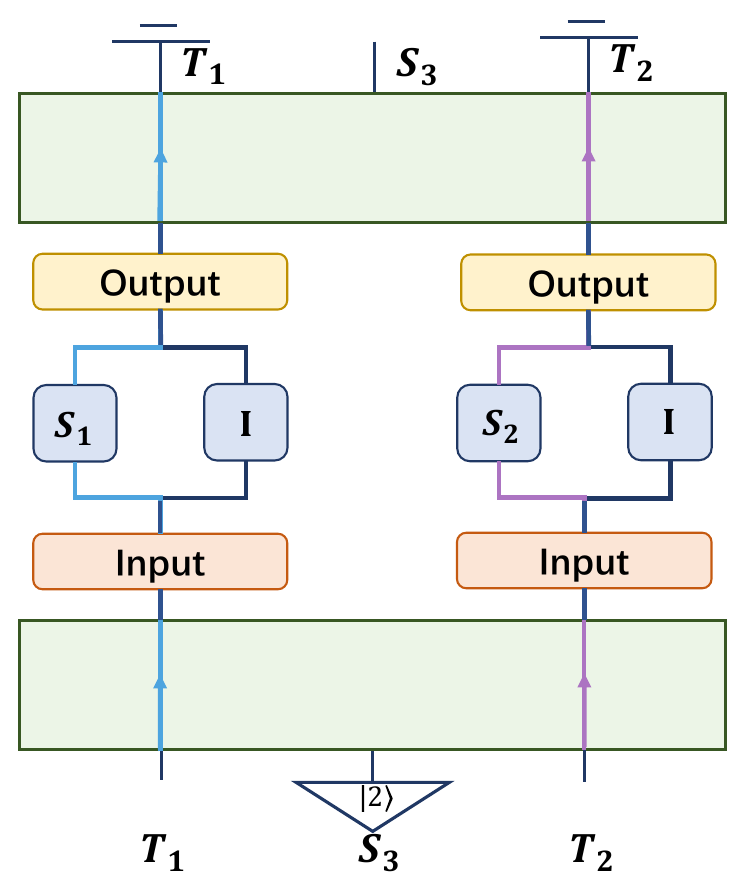}}
	\caption{Quantum PAR-SER switch. There are three different causal orders depending on the state of control qutrit $S_3$. The blue( or purple) line shows the process that target $T_1$( or $T_2$) goes through. The input channel (orange block) distributes $T_1,T_2$ to parties or identity channel in different cases, respectively. Finally, the output channel (yellow block) send them back to quantum PAR-SER switch.
 }
\label{fig:Q_PARswitch}
\end{figure}

When the state of $S_3$ collapses to $|0\rangle$, the switch enables unidirectional signaling from $S_1$ to $S_2$, i.e., $S_1 \preceq S_2$. In this scenario, $T_1$ is firstly sent to $S_1$ for a quantum operation. And then the PAR-SER switch distribute $T_1$ to $S_2$. while $T_2$ remains unchanged all the time. Finally, the target particles are discarded (see Fig.  ~\ref{fig:Q_PARswitch.AB}). When the control qutrit is in the state $|1\rangle$, the direction of the signaling channel on $T_1$ is reversed, while $T_2$ remains unchanged (see Fig.  ~\ref{fig:Q_PARswitch.BA}). With $S_3$ in state $|2\rangle$, signaling between parties is blocked. $S_1$ and $S_2$ interact independently with $T_1$ and $T_2$ in parallel (see Fig.  ~\ref{fig:Q_PARswitch.NS}).

\begin{figure}
    \centering
    \includegraphics[scale=0.45]{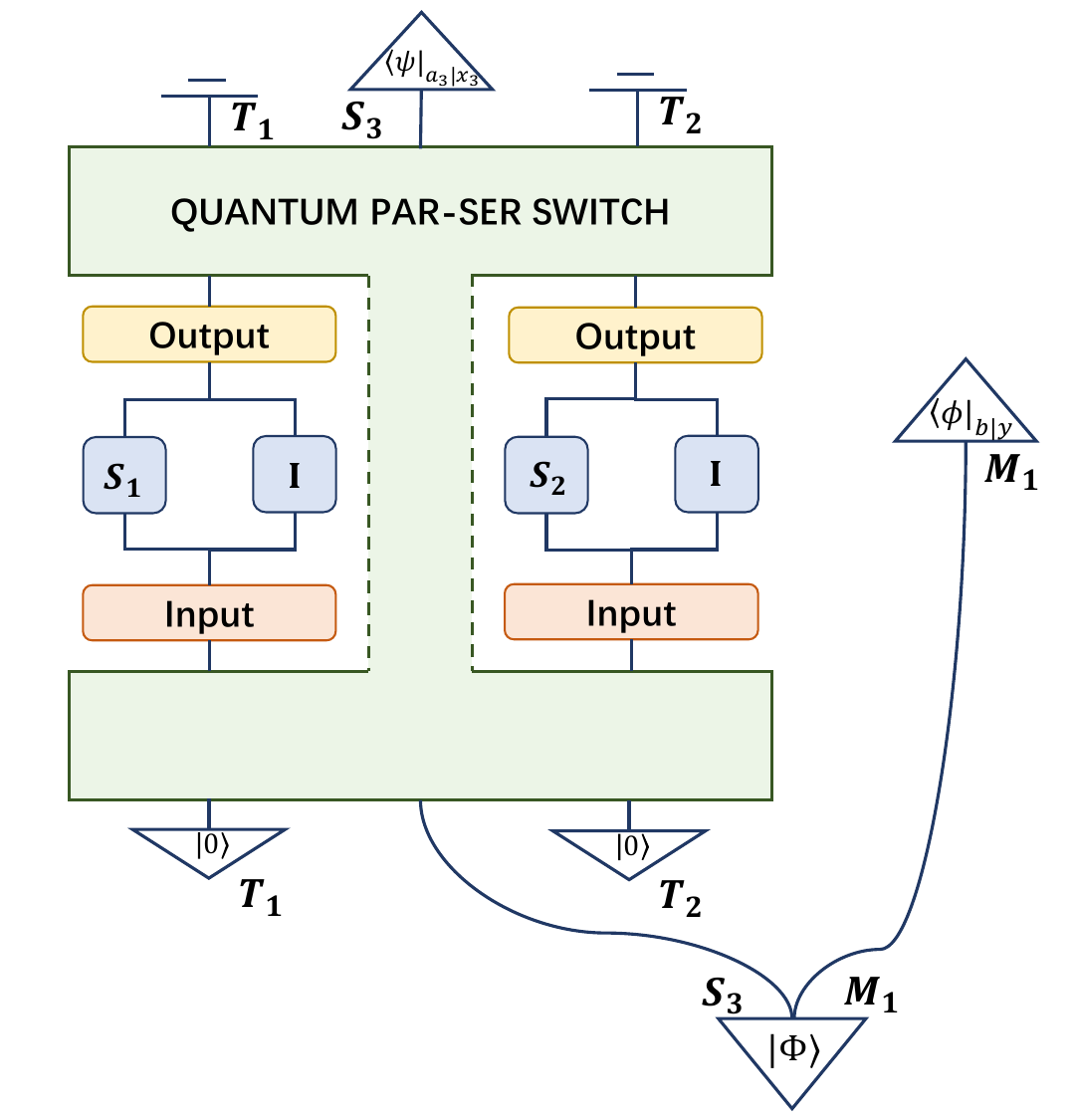}
    \caption{The quantum PAR-SER switch setup violating causal inequality. The control qutrit $S_3$ is prepared to be entangled with an auxiliary system $M_1$ in state $|\Phi\rangle$. Both of two target particle are prepared in state $|0\rangle$, then sent to the quantum PAR-SER switch. Finally, the corresponding measurement effects $\langle \psi |, \langle \phi |$ are conducted by $S_3$ and $M_1$ respectively, and two target particles are discarded.}
    \label{fig:Q_PARswitch_CER}
\end{figure}

In the process matrix framework, the identity channel is described by the projector onto the CJ vector~\cite{AraujoWitnessingcausalnonseparability2015}. Then the quantum PAR-SER switch is represented by a causally non-separable process matrix $W_{Q-PAR-SER}=|w\rangle\langle w|$, where $| w \rangle$ is defined as
\begin{equation}
    \begin{split}
        | w \rangle &= \frac{1}{\sqrt{3}} [|\psi\rangle^{x_1} |\mathbb{I}\rangle\rangle^{a_1x_2}|\mathbb{I}\rangle\rangle^{a_2T_{1I}}|\psi\rangle^{T_{2I}}|0\rangle^{x_3} 
        \\
        &+|\psi\rangle^{x_2} |\mathbb{I}\rangle\rangle^{a_2x_1}|\mathbb{I}\rangle\rangle^{a_1T_{1I}}|\psi\rangle^{T_{2I}}|1\rangle^{x_3} 
        \\
        &+ |\psi\rangle^{x_1} |\mathbb{I}\rangle\rangle^{a_1T_{1I}}|\psi\rangle^{x_2}|\mathbb{I}\rangle\rangle^{a_2T_{2I}}|2\rangle^{x_3}]
    \end{split}
\end{equation}
Here $T_{1I}$, $T_{2I}$ represent the input spaces of target qubits $T_1$ and $T_2$ respectively.
 
To experimentally certify the existence of causal nonseperability in this process, two target systems are initialized in the states $|0\rangle_{T_1}$ and $|0\rangle_{T_2}$. The control qutrit $S_3$ is maximally entangled with an auxiliary system $M_1$, resulting in the joint state $|\Phi\rangle_{S_3M_1}= \frac{1}{\sqrt{3}}(|00\rangle\langle00|+|11\rangle\langle11|+|22\rangle\langle22|)$. Inside each laboratory, party perform a set of quantum instruments defined by Kraus operators $|x_i\rangle\langle a_i|:\mathcal{H}_{T}\to \mathcal{H}_{T}$. When a target system $T_j(j\in \{1,2\})$  is received, $S_i(i\in\{1,2\})$ measures it and records the outcome as $a_i$.  $x_i$ is then encoded in the basis of the outgoing target qubit $T_j$, and send it away. Finally, the measurement outcomes for $S_3$ and $M_1$ are obtained using projective measurements $\langle\psi |_{a_3|x_3}:\mathcal{H}_{S_3} \to \mathds{R}$ and $\langle\phi |_{b|y}:\mathcal{H}_{M_1} \to \mathds{R}$, respectively (see Fig.  \ref{fig:Q_PARswitch_CER}). The probability of obtaining the joint outcomes $a_1,a_2,a_3,b$ given the settings $x_1,x_2,x_3,y$ is calculated using the Born rule, as shown in Eq. (\ref{Appdeq:p}).
\begin{equation}
    \begin{split}
       & p(\Vec{a}b|\Vec{x}y)=|\langle\psi |_{a_3|x_3} \otimes \langle\phi |_{b|y} (|0\rangle\langle0| \otimes |x_2\rangle\langle a_2|x_1\rangle\langle a_1|_{T_1}\otimes \mathbb{I}_{T_2}
       \\
       &+|1\rangle\langle1| \otimes |x_1\rangle\langle a_1|x_2\rangle\langle a_2|_{T_1}\otimes \mathbb{I}_{T_2}
        \\
        &+|2\rangle\langle2| \otimes |x_2\rangle\langle a_2|_{T_1}\otimes |x_1\rangle\langle a_1|_{T_2}) |\Phi\rangle_{S_3M_1}|0\rangle_{T_1}|0\rangle_{T_2}|^{2}
    \label{Appdeq:p}
    \end{split}
\end{equation}

During the experiment, when $ y = 0$ , we postselect on the process yielding the outcome $b = i$  (with  $i = 0, 1, 2$). It corresponds to the same correlations in the switch  when the control qutrit is in the state $|i\rangle_{S_3}$. As a result, the switch perfectly wins the corresponding game, ensuring that  Eq. (\ref{Appdeq:alphanota}) equals to $1$. The last term represents  an independent Bell test conducted by $S_3$ and $M_1$. When $S_3$ and $M_1$  employ the Bell operators described in ~\cite{SalavrakosBellInequalitiesTailored2017}, quantum mechanics predicts the maximal violation of the inequality $I_3$, i.e. $I_3(a_3,b|x_1=x_2=0,x_3,y) \leq 4$. Therefore, the inequality given in Eq. (\ref{Appdeq:inequa}) is violated.
\begin{equation}
\begin{split}
    &\alpha+\frac{1}{4(3+\sqrt{3})}I_3(a_3,b|x_1=x_2=0,x_3,y) 
    \\
    &=1+\frac{1}{3+\sqrt{3}} > \frac{7}{8}+\frac{1}{2\sqrt{3}}
\end{split}
\end{equation}
\section{Conclusions}

In this study, we bridge foundational research motivated by causality, specifically connecting GPTs with the process matrix framework. We demonstrate that in tripartite systems, the no-signaling principle is dual to classical processes, revealing a trade-off between them. By showing that the spanning vector sets of measurement spaces in box world and local theory are identical, we establish an equivalence between classical processes and measurement vectors in these frameworks, thereby providing an axiomatic definition of measurement space.

Furthermore, we identify that not all effects in the dual space are permissible, as some measurements require fine-tuning for causal interpretation. This observation allows us to refine our analysis to focus on deterministic classical processes. The study of causality in effects extends beyond box world and local theory. Importantly, certain effects within almost quantum correlations cannot be described as wirings~\cite{sainzAlmostQuantumCorrelationsViolate2018}, exploring the indefinite causal order of measurements in GPTs that pre, highlighting intriguing possibilities when exploring the indefinite causal order of measurements in GPTs that predict almost quantum or quantum correlations. This direction presents a promising avenue for future research.

We also explore logically consistent classical processes in 4-partite systems, presenting examples that do not manifest in tripartite cases. One such example is the introduction of a novel causal order termed the PAR-SER switch, which we certify in a quantum variant through an inequality. This switch allows parties to coherently control operations in both parallel and serial configurations. As the number of control states increases, the potential for more complex operational strategies grows, enabling richer interactions among the involved parties.

However, further investigation is needed into more intricate indefinite causal orders in 4-partite systems, particularly beyond simple self-loops. Recent experiments on non-classical causal orders, such as the quantum switch, have shown great promise for quantum information processing, including applications in quantum communication, computation, and metrology~\cite{ProcopioExperimentalsuperpositionorders2015a,ChiribellaPerfectdiscriminationnosignalling2012b,ZhaoQuantumMetrologyIndefinite2020a,RennerComputationalAdvantageQuantum2022,GuerinExponentialCommunicationComplexity2016,ChapeauNoisyquantumindefinitecausalorder2021,AraujoComputationalAdvantageQuantumControlled2014b,FeixQuantumsuperpositionorder2015,YinExperimentalsuperHeisenbergquantum2023}. Thus, future research will naturally focus on the practical implementation and potential advantages of the quantum PAR-SER switch, especially in these cutting-edge fields.

\bibliography{biblio}

%\appendix
%\clearpage

\end{document}